%% file: XeRecoil-prc.tex
\documentclass[twocolumn,superscriptaddress,prd,showkeys,showpacs,nofootinbib]{revtex4-1}

\usepackage{graphicx}
\usepackage[colorlinks = true,
            linkcolor = cyan,
            urlcolor  = blue,
            citecolor = red,
            anchorcolor = blue]{hyperref}\usepackage{color}
\usepackage{amssymb}
\usepackage[nointegrals]{wasysym}
\usepackage{amsthm}
\usepackage{amsmath}
\usepackage{textcomp}
\usepackage{mathtools}
\usepackage{subcaption}
\usepackage{multirow}

\usepackage{lineno}

\usepackage{comment}
\usepackage[separate-uncertainty,retain-explicit-plus,per-mode = symbol]{siunitx}
\usepackage{url}
\usepackage{array}
\newcolumntype{C}[1]{>{\centering\arraybackslash}p{#1}}

\input{definitions.tex}

\begin{document}

\title{Measurement of the ionization yield from nuclear recoils in liquid xenon between 0.3 -- 6 keV with single-ionization-electron sensitivity}

\author{B.G.~Lenardo} \affiliation{Physics Department, Stanford University, 382 Via Pueblo Mall, Stanford, CA 94305, USA} \affiliation{Lawrence Livermore National Laboratory, 7000 East Ave., Livermore, CA 94551, USA} \affiliation{University of California Davis, Department of Physics, One Shields Ave., Davis, CA 95616, USA}  
\author{J.~Xu} \email[Corresponding author, ] {xu12@llnl.gov}\affiliation{Lawrence Livermore National Laboratory, 7000 East Ave., Livermore, CA 94551, USA}  
\author{S.~Pereverzev}  \affiliation{Lawrence Livermore National Laboratory, 7000 East Ave., Livermore, CA 94551, USA}  
\author{O.A.~Akindele}  \affiliation{Lawrence Livermore National Laboratory, 7000 East Ave., Livermore, CA 94551, USA}  
\author{D.~Naim} \affiliation{University of California Davis, Department of Physics, One Shields Ave., Davis, CA 95616, USA} 
\author{J.~Kingston} \altaffiliation{Current address: The University of Chicago, Division of the Physical Sciences, 5801 South Ellis Avenue Chicago, Illinois 60637, USA} \affiliation{Lawrence Livermore National Laboratory, 7000 East Ave., Livermore, CA 94551, USA}  
\author{A.~Bernstein} \affiliation{Lawrence Livermore National Laboratory, 7000 East Ave., Livermore, CA 94551, USA}  
\author{K.~Kazkaz} \affiliation{Lawrence Livermore National Laboratory, 7000 East Ave., Livermore, CA 94551, USA}  
\author{M.~Tripathi} \affiliation{University of California Davis, Department of Physics, One Shields Ave., Davis, CA 95616, USA} 
\author{C.~Awe} \affiliation{Department of Physics, Duke University, and Triangle Universities Nuclear Laboratories, Durham, NC 27710}  
\author{L.~Li} \affiliation{Department of Physics, Duke University, and Triangle Universities Nuclear Laboratories, Durham, NC 27710}  
\author{J.~Runge} \affiliation{Department of Physics, Duke University, and Triangle Universities Nuclear Laboratories, Durham, NC 27710}  
\author{S.~Hedges} \affiliation{Department of Physics, Duke University, and Triangle Universities Nuclear Laboratories, Durham, NC 27710}  
\author{P.~An} \affiliation{Department of Physics, Duke University, and Triangle Universities Nuclear Laboratories, Durham, NC 27710}  
\author{P.~S.~Barbeau} \affiliation{Department of Physics, Duke University, and Triangle Universities Nuclear Laboratories, Durham, NC 27710}

\date{\today}

\begin{abstract}
Dual-phase xenon TPC detectors are a highly scalable and widely used technology to search for low-energy nuclear recoil signals from WIMP dark matter or coherent nuclear scattering of $\sim$MeV neutrinos. Such experiments expect to measure O(keV) ionization or scintillation signals from such sources.
However, at $\sim1\,$keV and below, the signal calibrations in liquid xenon carry large uncertainties that directly impact the assumed sensitivity of existing and future experiments.
In this work, we report a new measurement of the ionization yield of nuclear recoil signals in liquid xenon down to 0.3$\,$keV$\,\,$-- the lowest energy calibration reported to date -- at which energy the average event produces just 1.1~ionized~electrons. 
Between 2 and 6$\,$keV, our measurements agree with existing measurements, but significantly improve the precision. At lower energies, we observe a decreasing trend that deviates from simple extrapolations of existing data.
We also study the dependence of ionization yield on the applied drift field in liquid xenon between 220V/cm and 6240V/cm, allowing these measurements to apply to a broad range of current and proposed experiments with different operating parameters.

\end{abstract}


\maketitle

\section{Introduction}
\label{sec:alt_intro}

\input{alt_introduction.tex}

\section{Experimental setup}
\label{sec:setup}

\input{experimental_setup.tex}

\section{Data analysis}
\input{data_analysis.tex}

\section{Measurement of the ionization yield and its field dependence}

\input{results_full_table.tex}

\section{Discussion}
\input{discussion.tex}

\section{Conclusion}
\input{conclusion.tex}

\begin{acknowledgments}

 This project is supported by the U.S. Department of Energy (DOE) Office of Science, Office of High Energy Physics under Work Proposal Number SCW1508 and SCW1077 awarded to Lawrence Livermore National Laboratory (LLNL). 
 
 LLNL is operated by Lawrence Livermore National Security, LLC, for the DOE, National Nuclear Security Administration (NNSA) under Contract DE-AC52-07NA27344.
 Certain equipment used in this measurement was recycled from the LLNL DUS Second Time Around store. TUNL is operated as a DOE Center of Excellence under Grant Number DE-FG02-97ER41033.
 D.~Naim is supported by the DOE/NNSA under Award Number DE-NA0000979 through the Nuclear Science and Security Consortium. Duke University student support is provided by the DOE under Award Number DE-SC0014249.  
 
 We thank Sean Durham and Jesse Hamblen from LLNL, 
 and Dave Hemer, Keith Delong and Michael Irving from UC Davis for their technical support 
 on designing and constructing the xenon TPC detector. We also thank Vladimir~Mozin and Phillip~Kerr for their help during early tests, the TUNL scientific and technical staff for their assistance during setup and operation, and Prof.~John~Mattingly of North Carolina State University for lending us the liquid scintillator detectors used in this work. 
 Finally, we thank the members of the LUX and LZ collaborations for their constructive discussions on this measurement and possible implications.

\end{acknowledgments}

\bibliographystyle{apsrev}
\bibliography{biblio,BriansGlobal}

\end{document}

%% file: definitions.tex
\newcommand{\us}{$\mu$s}

%% file: alt_introduction.tex
Dual-phase xenon time projection chambers (TPCs) are widely used in fundamental physics for the detection of rare low-energy signals. In particular, they are currently used in several of the most sensitive searches for Weakly Interacting Massive Particles (WIMPs)~\cite{LUXRun4PRL,XENON1T_2018,PandaXSpinIndependent2016}, a class of hypothesized particles which provide a compelling solution to the dark matter problem. These experiments seek to detect ionization and scintillation produced in the liquid xenon target by the scattering of WIMPs with xenon nuclei. 
As a result of the low expected energy transfer for light WIMPs scattering on relatively heavy xenon nuclei, the sensitivity to WIMP masses below $\sim10$~GeV/$c^2$ can be greatly enhanced by lowering the nuclear recoil energy threshold below 3$\,$keV~\cite{LUXRun3Reanalysis}.
To reach the lowest-mass sensitivities, these experiments can drop the requirement for detecting scintillation light and operate in ``ionization-only" search mode, taking advantage of an $\mathcal{O}$(100\%) efficiency for the detection of ionization electrons (as opposed to $\mathcal{O}$(10\%) for scintillation photons) achieved by extracting the electrons into the gas phase and measuring a proportional electroluminescence signal~\cite{XENON10_S2only}.

In addition to WIMP dark matter, low energy nuclear recoils can be produced in such a detector via coherent elastic neutrino-nucleus scattering (CE$\nu$NS). This interaction was recently measured for the first time in a CsI target~\cite{COHERENT2017_CENNS}, and is a subject of interest as a new probe of standard model and beyond-the-standard-model physics~\cite{DuttaZprime2016,DuttaSterile2016,Anderson2012,COHERENT_NuChargeRadii,Canas_CoherentWeakMixingAngle2018}. It further provides a channel for the flavor-independent detection of solar and supernova neutrinos in the next generation of xenon-based WIMP dark matter detectors~\cite{LZSensitivityPaper,Lang2016_SupernovaXe}. There is also interest in taking advantage of the enhanced cross section of CE$\nu$NS relative to the inverse beta decay (IBD) reaction, which is widely used in experiments detecting reactor antineutrinos, to construct smaller-footprint neutrino detectors for nuclear nonproliferation applications~\cite{HagmannBernstein}. As in the case of low-mass WIMPs, the expected signal from each of these sources is a steeply falling spectrum of nuclear recoils with energies of $\mathcal{O}$(1$\,$keV). 

To accurately calculate expected rates and analyze experimental data, it is crucial to have a precise calibration of the energy scale for nuclear recoils at the energies of interest. The ionization yield, or electrons produced per unit energy, is particularly important due to the previously-stated low-threshold capabilities of measurements using the ionization channel. There is an inherent challenge, however, in that the stopping of nuclei is distinct from electronic recoils; a large fraction of energy in nuclear recoils is lost as heat. Therefore, while electronic recoil calibrations can be performed using standard x-ray, $\beta$ or $\gamma$-ray sources, nuclear recoil signals need to be calibrated independently. 

Previous efforts to measure the ionization yield in liquid xenon have used one of two techniques. The first use broad-spectrum neutron sources, such as AmBe or $^{252}$Cf, and fit the measured spectra to extract the ionization yield as a continuous function of energy~\cite{ColumbiaQy2006,HornQy2011,SorensenQy2009,Xenon100Qy2013}. The second use fixed-angle scattering of monoenergetic neutrons, in which the angle of the scattered neutron is either measured in-situ or by secondary detectors to constrain the recoil energy~\cite{ManzurQy2010,LUXDD,Aprile2018_Yields}. Of these measurements, only those in Refs.~\cite{LUXDD}~and~\cite{SorensenQy2009} extend below 3$\,$keV, and only the former goes below 1$\,$keV, with $\sim$30\% uncertainties at the lowest energy point (0.7$\,$keV). These results allow robust searches for WIMPs above $\sim$10$\,$GeV/c$^2$, but for signals below this value the uncertainties increase quickly due to the lack of precise data at lower energies. Searches for the coherent neutrino scattering signals discussed above face similar difficulties. A precise calibration of the low energy ionization yield is therefore needed for understanding the absolute limits of liquid xenon-based detectors, and their sensitivities to new physics in these various applications.

In this work, we use a pulsed, monoenergetic neutron source at the Triangle Universities Nuclear Laboratory (TUNL) to produce nuclear recoils in liquid xenon with energies from 0.3 to 6~keV. Scattered neutrons are tagged by an array of ten liquid scintillator detectors placed at fixed angles, kinematically constraining the recoil energies for events detected in each channel. Using time-of-flight~(TOF) analysis and pulse shape discrimination (PSD) in the tagging detectors to reduce backgrounds, we are able to measure ionization signals down to the single-electron level with $\mathcal{O}$(100\%) efficiency. This allows us to robustly calibrate liquid xenon ionization production at sub-keV energies, and provides the highest precision measurement of the low-energy ionization yield from nuclear recoils to date. We further measure the dependence on the applied electric field, allowing our calibration to apply generally to liquid-xenon-based ionization detectors with a broad array of operating conditions. We measure for the first time a statistically significant dependence of the charge yield on the applied drift field between 1$-$6~keV, contrary to recent results between 5$-$14~keV~\cite{Aprile2018_Yields}. We also measure for the first time nuclear recoil ionization signals of $\sim$1 electron, approaching the fundamental limit of nuclear recoil detection in liquid xenon. Our results allow reliable calculations of the sensitivity of present and future xenon-based experiments searching for new physics via nuclear recoils at the few-electron level.

%% file: experimental_setup.tex
\subsection{Liquid xenon detector}
\label{subsec:lxe_detector}
\begin{figure}[t!]
\centering
\includegraphics[width=.46\textwidth]{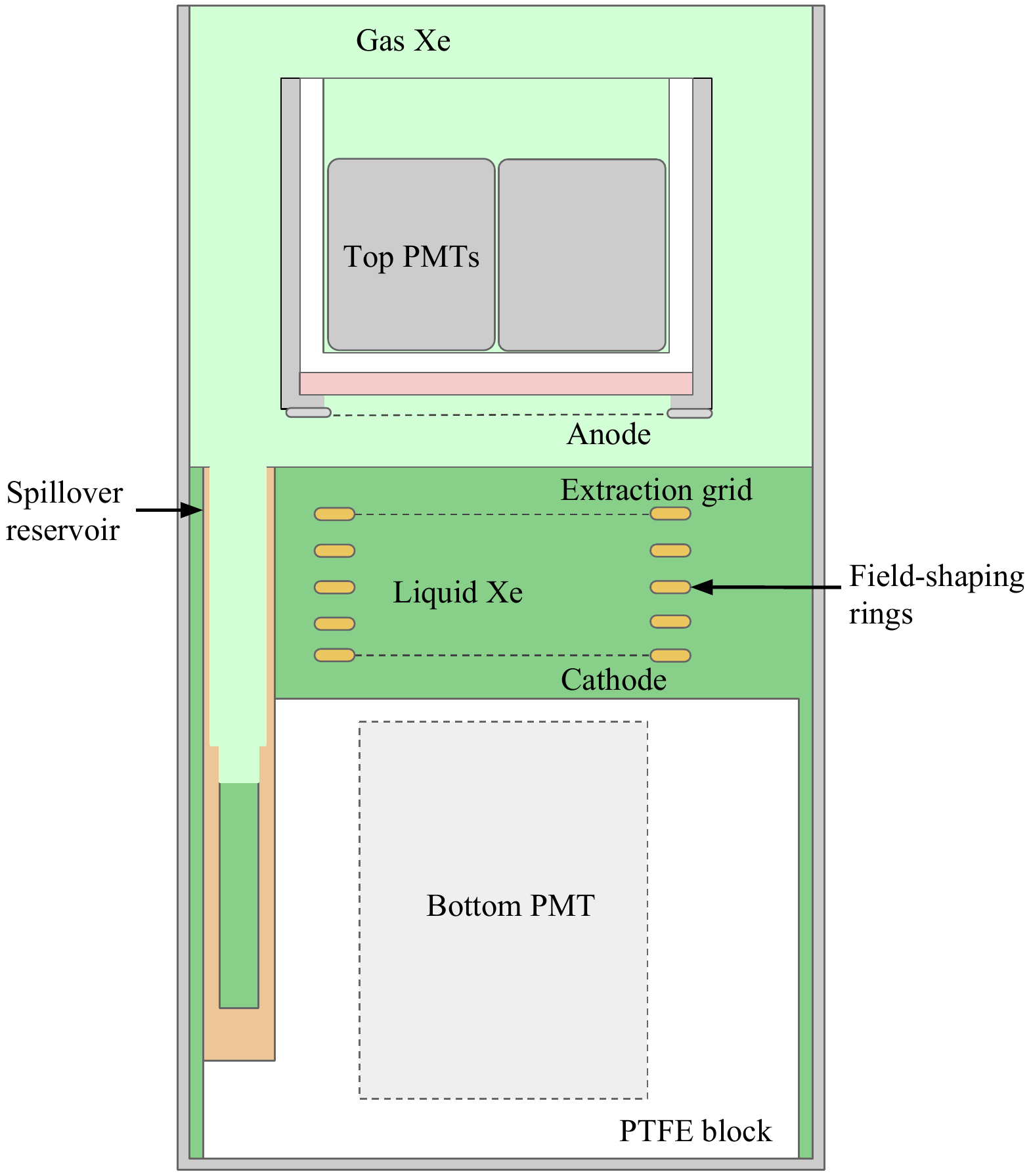} 
\caption{Schematic of the dual-phase xenon detector. Details are given in Section~\ref{sec:setup}.}
\label{fig:detector_schematic}
\end{figure}

For these measurements we designed a dedicated, compact dual-phase xenon TPC. A schematic of the detector internals is shown in Fig.~\ref{fig:detector_schematic}. The active volume consists of 140~g of liquid xenon ($\diameter$5~cm$\times$2.5~cm), surrounded by an electric field-shaping cage made of copper rings connected through 1$\,\text{G}\Omega$ resistors.
Electrons released by ionizing radiation in the active volume, under the influence of the applied electric field, drift across the bulk to the liquid surface and are extracted into the gas, where they produce an electroluminescence (EL) signal proportional to the number of extracted electrons. The EL light is collected by five photomultiplier tubes:
an array of four Hamamatsu R8520-406 PMTs 
measures the light from above, giving horizontal position information,
and one Hamamatsu R8778 PMT 
is immersed in the liquid below the active xenon to increase the light collection efficiency. The PMTs also detect primary scintillation originating in the liquid volume, which is distinguished from the EL signals by timing information (scintillation pulses are O(10$\,$ns) wide, compared to O(1$\,\mu$s) for EL signals) and ignored in this analysis.

Three hexagonal mesh grids (90\% optically transparent at normal incidence) can be biased to independently apply electric fields across the target volume and the liquid-gas interface.
In normal operation, two custom feedthroughs provide negative high voltage (HV) to the extraction grid and the cathode, 
while the anode remains grounded.
The position of the liquid-gas interface is fixed by the height of a PEEK spillover reservoir; following the treatment in~\cite{PIXeY2018_EEE}, we calculate a $0.5\pm0.5\,$mm head of liquid above the edge of the reservoir during regular operation. Accounting for these effects, the liquid level in the detector is $7.5\pm0.5\,$mm above the extraction grid and $5.9\pm0.5\,$mm below the anode. 
Throughout the measurements presented here, a bias voltage of 12.0~kV was applied to the extraction grid. A full 3D COMSOL simulation was performed to calculate the resulting electric fields. In the extraction region, we calculate an electric field of 11.7 kV/cm in the gas and 6.24$\,$kV/cm below the liquid surface. 
This corresponds to an efficiency of $\sim95$~\% for extracting electrons from the liquid into the gas~\cite{LLNL_EEE2019}. The voltage applied to the cathode was varied between 12.5$\,$kV and 17.0$\,$kV to vary the drift field in the active volume.

Xenon is continuously circulated in a closed loop through a hot SAES MonoTorr getter to remove electronegative impurities. The liquid is drawn from the spillover reservoir into a heat exchanger, where it is evaporated and pumped through the purifier. It then recirculates back through the heat exchanger and is re-condensed above the xenon volume. The liquid drips down through a PTFE tube to the bottom of the detector vessel. The purification mass flow rate was set at 1.2 standard liter per minute, 
turning over all $\sim$1.5~kg of xenon in the system in 4~hours. 
At this circulation rate, the liquid level in the reservoir remained below the active detector volume for the duration of data taking.

\subsection{TUNL neutron facility}
\label{subsec:neutron_beam}

Neutron scattering measurements were carried out at the 10MV tandem accelerator at TUNL.
Neutrons were generated via the the Li(p,n)Be reaction on a tantalum-backed LiF target of thickness 200$\mu$g/cm$^2$. The mean energy and spread for the forward-going neutrons was measured in-situ using a time-of-flight (TOF) technique, finding a mean energy of $579\pm3\,$keV with a 1-$\sigma$ spread of $10\,$keV. These measurements were consistent with the distribution calculated analytically from proton energy dissipation in the LiF film. Pulses produced by the accelerator were $\sim$1$\,$ns long with an intrinsic period of 400~ns. We kept one of every eight pulses to reduce random coincidence backgrounds, giving an interval of 3.2$\,\mu$s between pulses. The proton-on-target time was measured with an inductive pickup circuit, hereafter referred to as the beam pulse monitor (BPM), and the overall beam current was measured directly on the target. We measured a current between 50 and 90$\,$nA for the duration of our measurement campaign. 

\begin{figure}[h!]
\centering
\includegraphics[width=.49\textwidth]{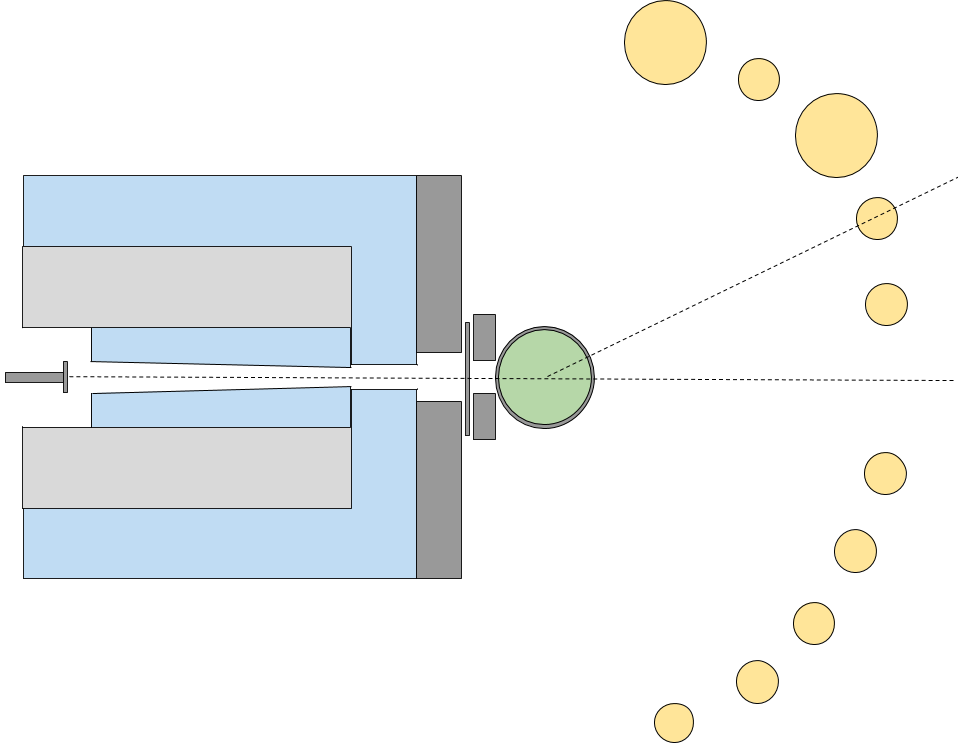} 
\caption{Diagram of experimental setup. The collimator and shielding are color-coded to show the borated polyethylene (blue), high-density polyethlyene (light grey), and lead (dark grey). The xenon TPC is shown in green, and the liquid scintillator backing detectors are shown in yellow.}
\label{fig:setup}
\end{figure}

Neutrons produced at the target were collimated at zero degrees using a mixture of borated polyethylene (BPE) and high density polyethylene (HDPE), illustrated in Figure~\ref{fig:setup}. The full shielding had an outer dimension of 56$\,$cm$\times$56$\,$cm$\times$55$\,$cm; 
a central hole, 3.8$\,$cm near the target tapering to 2.2$\,$cm near at the outlet, allowed neutrons at zero degrees to pass unimpeded. 
Gamma rays from neutron capture in the collimator were attenuated with 10~cm of lead shielding between the collimator and the xenon detector, with an opening of $\sim$7$\,$cm$\times \sim$5$\,$cm in the center to permit the passage of the collimated neutrons. To further attenuate capture gamma rays passing through the beam opening, two additional lead bricks ($5$~cm thickness) were added to reduce the horizontal opening in the shielding to 2.7~cm, and a lead sheet of 6~mm thickness was placed completely across the opening. The former stops both B-capture and H-capture gammas (478$\,$keV and 2.2$\,$MeV, respectively) originating throughout the shielding, while the latter is intended to specifically attenuate the B-capture gamma rays originating near the tapered beam pipe. The neutron mean free path in lead at 579$\,$keV is approximately 3$\,$cm, so the 6$\,$mm sheet does not significantly attenuate the neutron beam. The xenon detector itself was wrapped with a 6$\,$mm-thick lead sheet to attenuate radiation from the surrounding environment, with cutouts in both the front and the rear to allow collimated neutrons that scatter in the xenon volume to pass through a minimal amount of material.

The xenon detector was positioned so the active volume was centered at the same height as the neutron beam to accept on-beam neutrons. Incoming neutrons passed through the center of the gas-filled portion of the spillover reservoir (wall thickness of 0.6~mm) to further reduce scattering in passive materials. 

\subsection{Neutron tagging and data acquisition}
\label{subsec:daq}

Neutrons exiting the xenon detector at specific angles were tagged by an array of ten liquid scintillator (LS) detectors attached to an aluminum frame and held fixed at the beam height.\footnote{Two LS detectors with larger dimensions had their centers placed slightly above the beam height. This effect is included in the following analysis} Eight of the ten neutron detectors used cells of EJ-309 scintillator (5~cm-diameter by 5~cm-height active volume), while the  
the other two used cells of EJ-301 scintillator (10~cm-diameter by 7.6~cm-height active volume). To cross-calibrate the LS detectors, we varied their bias voltages to align the Compton edge of 662$\,$keV $\gamma$-rays from a $^{137}$Cs source. 
All of the detectors demonstrated excellent PSD capabilities for neutrons at 579~keV. 
To reduce spurious triggers from environmental gamma rays, each LS detector was wrapped with 2-3$\,$mm of lead shielding to reduce the background rate due to low energy gamma rays. 
The LS detectors were positioned at angles between 15 degrees and 70 degrees with respect to the incoming beam,
corresponding to xenon recoil energies between 0.3$\,$keV and 6$\,$keV for single-scatter neutron events. 

Data acquisition was triggered by a two-fold coincidence between the LS detector array and the xenon TPC. 
LS pulses above a preset threshold ($\lesssim$1/3 of the observable neutron kinetic energy) in any of the 10 LS backing detectors 
generated a gate of 25$\,\mu$s (40$\,\mu$s for a subset of the acquired data).
A valid TPC trigger in this window triggered the acquisition of waveforms from all detector channels. 
The TPC trigger required 3-fold coincidence among the top four Xe PMTs, with hardware thresholds for each channel set below the single photoelectron level.
This was verified to maintain close to 100\% efficiency for EL pulses down to single extracted electrons; we evaluate it in Section~\ref{subsec:calibration}. In addition, we exerted a 100~\us\ trigger holdoff after each accepted trigger and a 3~ms trigger veto following high energy events in the xenon TPC to reject correlated ionization backgrounds. 
This veto algorithm resulted in a deadtime of $\sim$30\% in the experiment. 

For each valid trigger, we digitize 16 waveforms: 5 from the Xe TPC, 
10 from the LS detector array, 
and 1 from the BPM. 
Each waveform has 32$\,\mu$s pre-trigger 8$\,\mu$s post-trigger (for the data sets that used a 40$\,\mu$s coincidence window, we digitized 50$\,\mu$s starting from 42$\,\mu$s before the trigger time).
The digitizer used in this experiment is a Struck SIS3316, 
with a timing resolution of 4ns and a digitizing precision of 14 bit (2V input range). 

Data were acquired over the course of six days in 35 separate acquisitions, each of which contained between one and four hours of livetime. The measurement campaign was split into three approximately equal segments, in which a different voltage was applied to the cathode to study drift-field-dependence of ionization production in the liquid. We applied 12.5~kV, 13.6~kV, and 17.0~kV. The associated field was calculated using the 3D COMSOL simulation, giving 220$\,$V/cm, 550$\,$V/cm, and 2200$\,$V/cm, respectively. The liquid volume above the extraction grid provided a fourth drift field dataset at 6240$\,$V/cm when integrated over the entire run. The difference in position for these latter events resulted in a slightly different recoil energy distribution; this is taken into account in the following analysis.

%% file: data_analysis.tex
We carry out our analysis using the pulses identified in the digitized waveforms after baseline subtraction and gain calibration. More than 100 event waveforms from all datasets were hand-scanned to ensure that pulse finding parameters are optimized and inefficiencies are negligible.

\subsection{Xenon detector calibration}
\label{subsec:calibration}
For the Xe TPC PMT signals, we use an adaptive baseline algorithm to separate the fast pulses from the much slower variations in the baseline noise, 
and zero-suppress waveform regions where no pulses were recorded.
The gain of each PMT in the Xe TPC is measured in situ using isolated spikes in the waveforms that are $>5\,\mu$s away from any high-energy pulses. The single-photoelectron peak is fitted by a gaussian function to extract the response of the PMT in units of ADC counts per photoelectron.

This procedure is repeated for each of the 35 acquisitions independently. The waveforms in each channel are then scaled to the acquisition-specific units of photoelectrons (PE), correcting for any time-dependence in the PMT gains. The calibrated waveforms from all five PMTs are added together to form a summed Xe TPC waveform that is used for pulse finding and pulse quantity evaluation. 

Four pulse quantities are used in this analysis. The integrated pulse area, defined as the integral of the summed Xe TPC pulse in units of photoelectrons, gives a measure of the charge extracted from the liquid. A pulse width parameter, defined as the 50\%-integral time minus the 10\%-integral time, is used to separate electroluminescence pulses from scintillation signals originating in the liquid. 
A top-bottom asymmetry parameter is defined as ($A_T - A_B$)/($A_T+A_B$), where $A_T$ is the summed pulse area in the top array and $A_B$ is the pulse area in the bottom PMT; this can also be used to ensure that pulses are consistent with electroluminescence signals. Finally, the $x$- and $y$-positions of electroluminescence pulses are calculated using a log-ratio technique, defined as
    \begin{align*}
    \begin{aligned}
        x &= \ln{(A_{x_-}/A_{x_+})} \\
        y &= \ln{(A_{y_-}/A_{y_+})} 
    \end{aligned}
    \end{align*}
where $A_{x_-}$, $A_{x_+}$, $A_{y_-}$ and $A_{y_+}$ are the pulse areas summed across two top PMTs in the positive and negative $x$- and $y$-dimensions, respectively. The log-ratio parameter is approximately linear within $\sim2\,$cm of the center of the detector, and we use it to select only pulses with $r\leq1.3\,$cm for further analysis.

We next calibrate the response of the detector to single electrons (SEs) extracted from the liquid. We select XeTPC pulses more than 3$\,\mu$s after the triggered pulse, to avoid bias from triggered events. The resulting spectrum shows a clear peak at $\sim$60~PE. To extract the average SE size ($\mu_{_{SE}}$) and width ($\sigma_{_{SE}}$), we fit this spectrum to the sum of three normal distributions, representing 1-, 2-, and 3-electron events; the mean and width of the $n$-electron distributions are constrained to be $n\times\mu_{_{SE}}$ and $\sqrt{n}\times\sigma_{_{SE}}$ respectively. The fit gives $\mu_{_{SE}} = 56.5\pm0.3$~PE/e$^-$ and $\sigma_{_{SE}} = 11.6\pm0.3$~PE/e$^-$ for the full measurement campaign. We use this value to calibrate our spectra throughout this analysis. The uncertainty of $0.5\%$ is an overall scaling error, but is smaller than other sources of uncertainty and therefore does not significantly contribute to the final error.

\begin{figure}[h!]
\centering
\includegraphics[width=.46\textwidth]{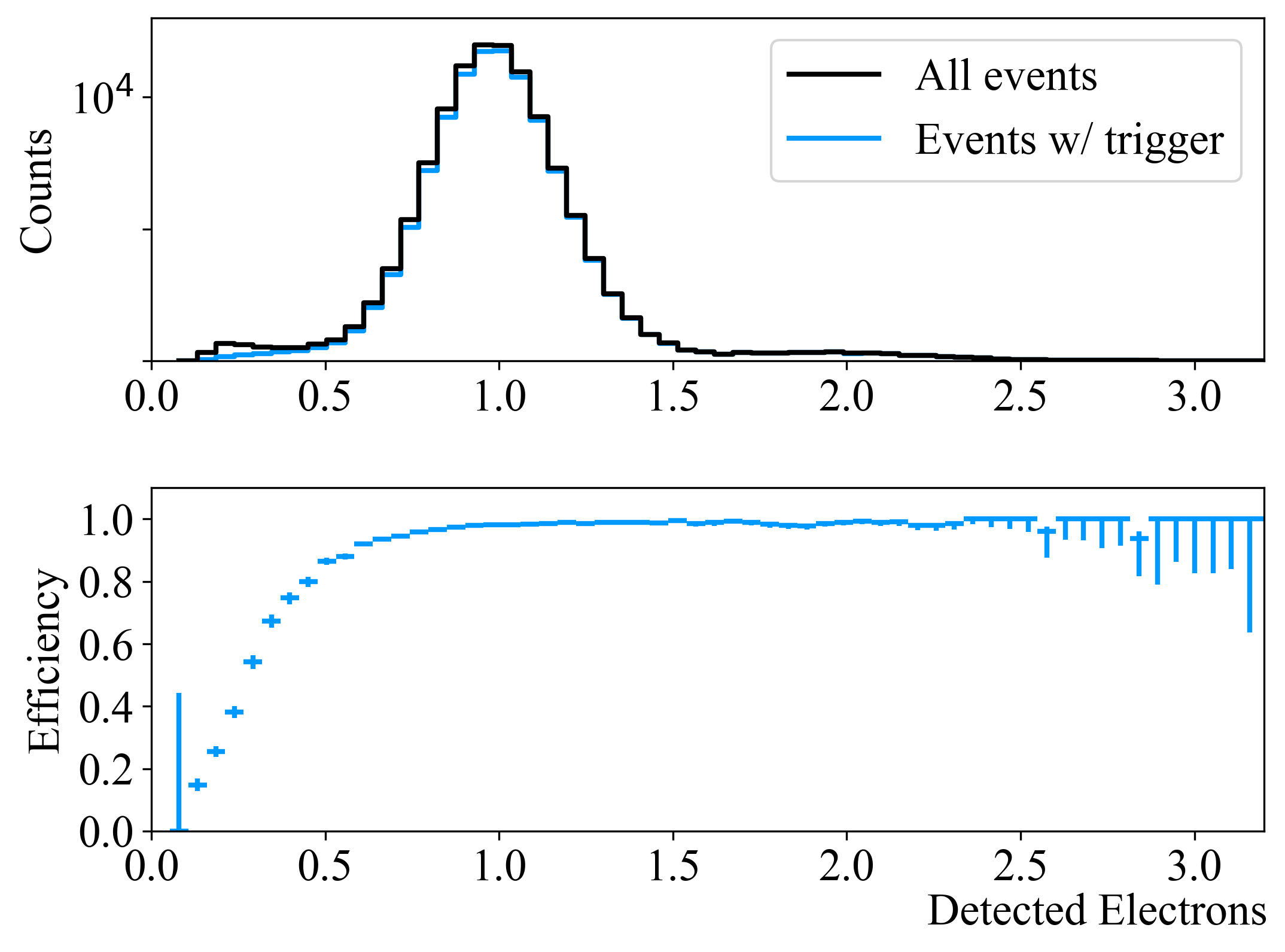} 
\caption{Efficiency of the three-fold coincidence trigger in the xenon TPC. The top shows the pulse area distribution (in units of PE) for events in a dedicated trigger evaluation dataset, as described in the text. The bottom shows the ratio of events that trigger the coincidence logic to the total number, giving the trigger efficiency in the xenon TPC.}
\label{fig:trigger_eff}
\end{figure}

We evaluate the Xe TPC trigger efficiency in situ using dedicated datasets in which the Xe TPC 3-fold coincidence signal was digitized directly. During these acquisitions, the DAQ was triggered not by the LS/TPC coincidence trigger, but by a discriminator which set a minimum-size threshold on the XeTPC pulses of $\sim$100~electrons. The efficiency of the xenon TPC coincidence signal was evaluated by selecting pulses in these events that occurred at least $5$~$\mu$s after the large pulse triggering the event. The fraction of these pulses that generated a xenon TPC logic signal gave us the coincidence trigger efficiency as a function of pulse area, and is shown in Figure~\ref{fig:trigger_eff}. The analysis confirms that we have $\sim$100\% efficiency for triggering on single extracted electrons.

\subsection{Neutron scattering data analysis}
For the LS neutron detectors, 
we use a simple flat-baseline algorithm and do not apply zero suppression due to the low PMT gains (single photoelectron pulse amplitudes in these detectors are typically below the baseline fluctuation). This method produces excellent PSD power and reasonable energy accuracy.

\begin{figure}[h!]
\centering
\includegraphics[width=.46\textwidth]{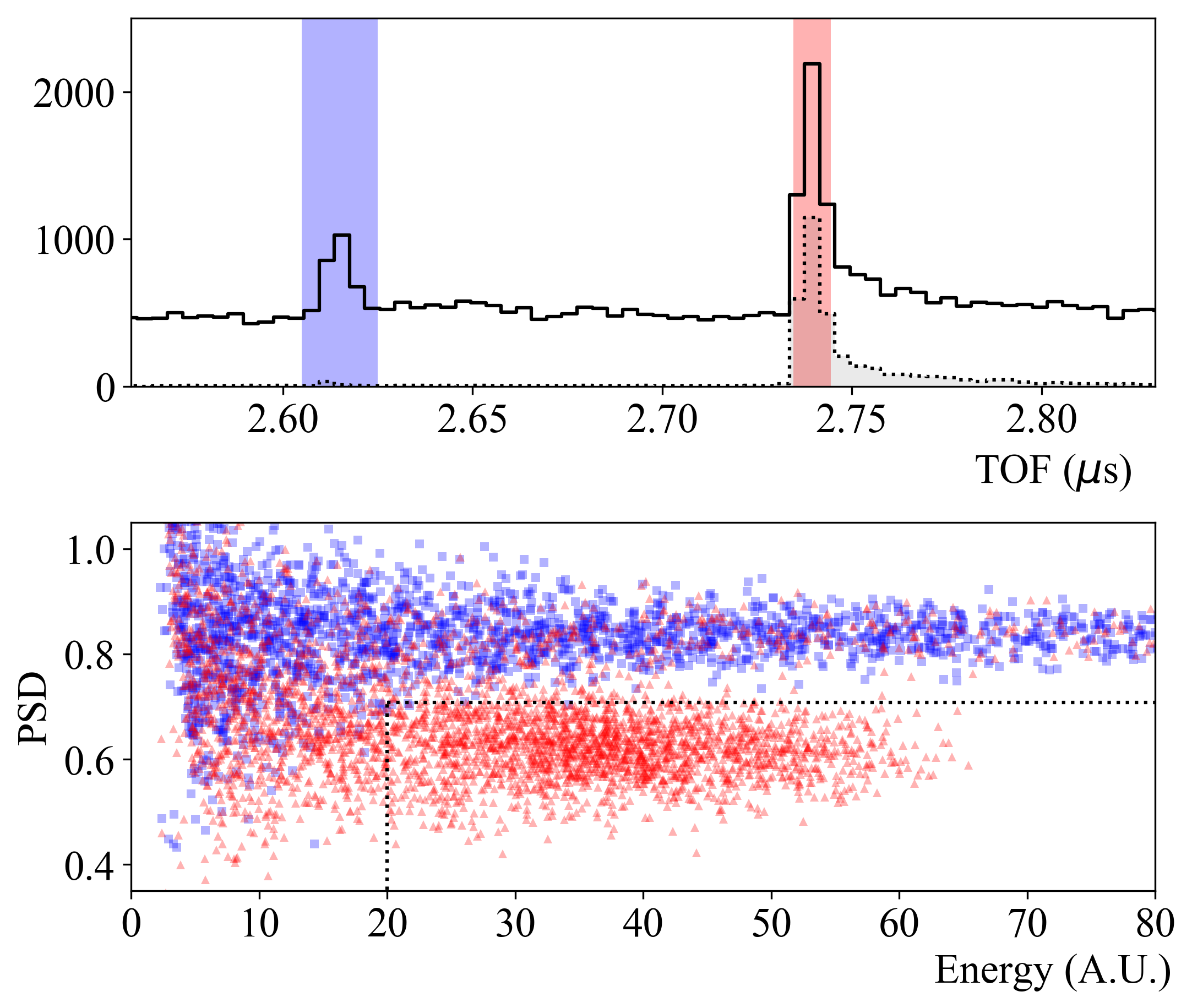} 
\caption{Time-of-flight (TOF) distribution (top) and the pulse shape discrimination parameter distribution (bottom). The neutron event selection cut in the TOF spectrum is shown by the red shaded region, and the corresponding PSD values of these events are shown by the red triangles in the bottom. For comparison, we also show events in the TOF gamma peak in blue. The PSD neutron selection cut is shown by the dotted grey line; the TOF distribution of all events passing this PSD cut is shown by the dotted grey histogram in the TOF spectrum.}
\label{fig:tof_PSD}
\end{figure}

Single-scatter-like neutron events are identified first by their time of flight, defined in our analysis as the difference between the LS pulse and the preceding BPM signal. To avoid any ambiguity, we discard events with signals in more than one LS detector. The small energy transfer from neutron elastic scattering on xenon guarantees that the single-scattered neutrons retain approximately their original kinetic energy, and therefore arrive at the LS detectors at a well-defined time relative to the beam pulse. The TOF can therefore be used to substantially reduce backgrounds from beam-related gammas, random coincidences, and multiple-scattering neutrons while retaining a high efficiency for the desired single-scatter events.
An example TOF distribution is shown in Figure~\ref{fig:tof_PSD} (top), 
where the beam-correlated gamma peak is observed at 2.61$\,\mu$s, the beam-correlated neutron peak is observed at 2.74$\,\mu$s, and the uncorrelated random coincidence backgrounds create a flat spectrum across all TOF values.
The absolute time scale in the TOF values is set by an unknown delay between the proton-on-target time and the BPM time, 
which can be subtracted off using the position of the gamma peak but is irrelevant for our analysis.

We select only events within the full width half maximum (FWHM) of the neutron TOF peak for further analysis.

We additionally select neutron events using the PSD capabilities in the LS backing detectors. The PSD parameter is defined as the fraction of the pulse area contained in the first 20$\,$ns after the half-maximum point on the rising edge of the LS pulse. Figure~\ref{fig:tof_PSD} (bottom) shows an example PSD distribution as a function of energy, with different selections in TOF. 
The band at PSD$\,\approx\,$0.85 is clearly associated with gamma rays, while the population at PSD$\,\approx\,$0.6 is associated with neutrons. We apply two cuts in this parameter space, shown by the dotted lines in Figure~\ref{fig:tof_PSD} (bottom).
These cuts select LS pulses with a PSD value no higher than 1.5-$\sigma$ from the peak of the neutron distribution and with recorded energies above half of the neutron peak in energy (at $\sim$40$\,$a.u.). The latter cut is imposed because of the deteriorating PSD capability at low energies. 
A large fraction of the neutrons deposit their full energy in the backing LS detectors, giving us a large sample of clean neutron scattering events that pass these two selection criteria.

For events passing the neutron-selection cuts, the first EL pulse recorded in the Xe TPC following the LS pulse is taken as the the neutron-induced Xe recoil signal. 
If the first pulse in the TPC arrives before the LS pulse, the whole event is discarded as a background coincidence. 
In cases when multiple TPC pulses are observed in the pre-trigger region, 
the detector is considered to be in a background-dominated state
and all events in the following 3$\,$ms are discarded via a software veto. 
For an ionization pulse to be selected for the following analysis, 
we also require its horizontal position to be within 1.3$\,$cm from the center of the TPC. 
It has been pointed out that there is an ambiguity in position reconstruction for detectors using a four-PMT setup~\cite{StephensonMIXPaper}. 
To address this, we further impose a top-bottom asymmetry cut. For events near the perimeter of the xenon volume, the enclosure surrounding the top PMTs occludes and reflects a larger fraction of light that would otherwise be recorded in the top array, leading to a smaller top-bottom asymmetry. This cut therefore rejects events near the perimeter of the TPC that may be mis-reconstructed into the center.

\begin{figure}[h!]
\centering
\includegraphics[width=.48\textwidth]{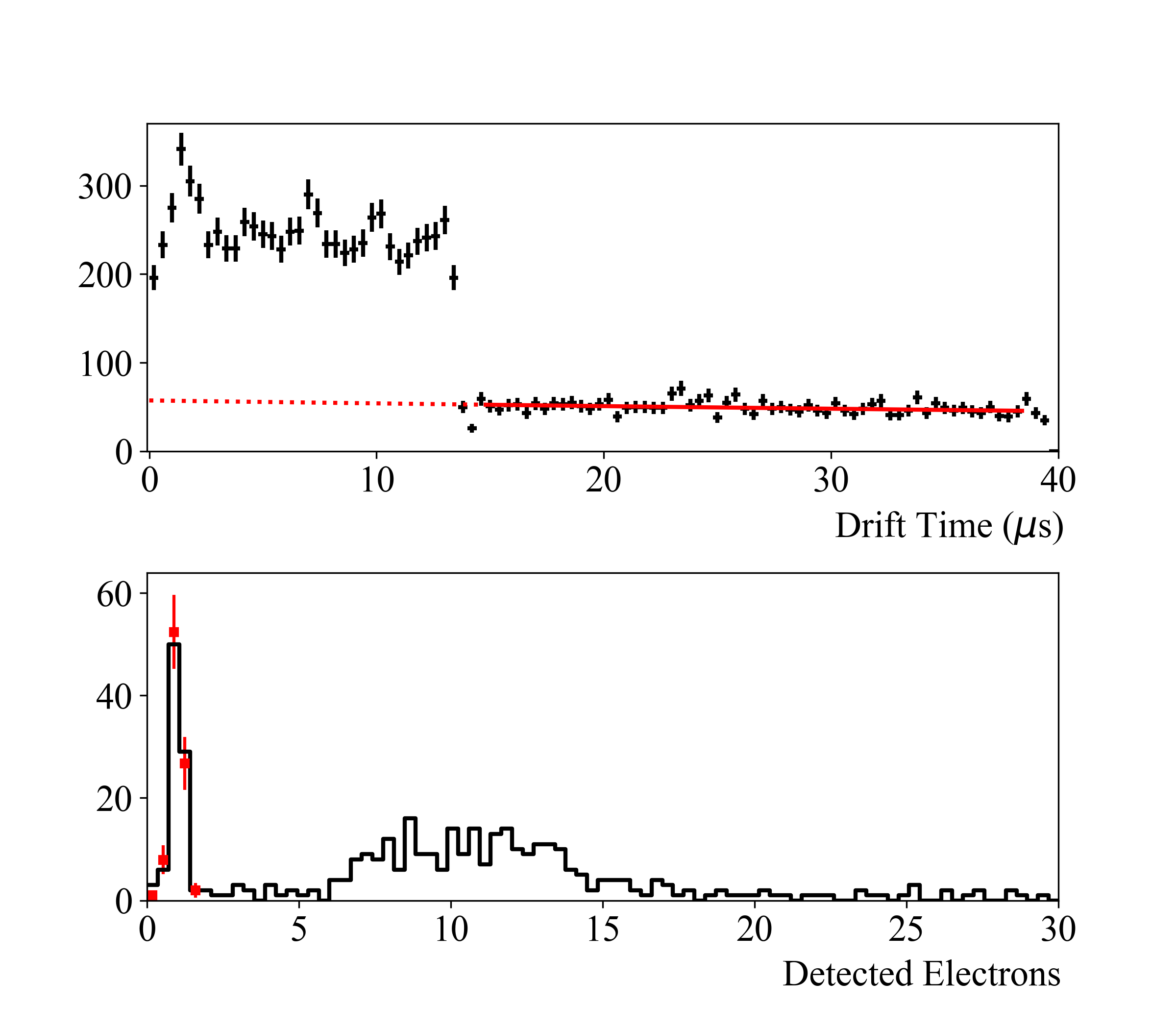} 
\caption{Coincidence data from Detector 6 (1.61~keV recoils) taken at a drift field of 2200~V/cm. The drift time distribution (top) shows a clear cutoff at the maximum drift time in the xenon TPC ($\sim$13~$\mu$s). Events beyond this value are from random coincidence triggers. 
The detected ionization distribution with a drift time cut applied is shown in the bottom (black curve). The background distribution (red points) is measured at drift times beyond 13$\,\mu$s and is scaled to an expected background rate using an exponential fit (red line, top) extrapolated to short drift times. Backgrounds are subtracted in all spectra before performing the charge yield analysis.}
\label{fig:dt}
\end{figure}

The drift time for ionization signals between the interaction point and the detected EL pulse is reconstructed by subtracting the LS pulse time from the time of the pulse in the xenon TPC. 
The black points in Figure~\ref{fig:dt} (top) show the distribution of the drift times in the Xe TPC for the datasets with a 40\us\ coincidence window. 
Events with a drift time smaller than 3~\us\ occur above the extraction grid (extraction volume), 
and events with drift times between 3 and 13~\us\ occur between the cathode and the extraction grids (target volume). 
Events with drift times above 13~\us\ are attributed to random coincidence between background ionization in the Xe TPC and LS pulses. The drift time is used in both background subtraction and ionization detection efficiency calculations, which are described in the following two sections.

An example ionization spectrum for BD channel 6 (1.6$\,$keV nuclear recoils), after all cuts have been applied, is shown in Figure~\ref{fig:dt} (bottom). The single-scatter nuclear recoil peak is clearly visible at $\sim$10~detected~electrons. There is also a continuum background due to neutron multiple-scattering events, as well as background peak of single-detected-electron signals which from random coincidence signals. 

\subsection{Random-coincidence background subtraction}
\label{subsec:background_subtraction}

Random coincidence backgrounds are subtracted from the final ionization spectra by directly subtracting the ionization spectra for events with drift times $>13\,\mu$s, scaled to the number of events expected with drift times consistent with drift-volume scattering events. This procedure is illustrated by the red curves in Figure~\ref{fig:dt}. The scaling is accomplished by fitting an exponential function to the drift time distribution at DT$>\,13\,\mu$s and extrapolating the event rate into the $3\mu\,$s$\,<\,$DT$\,<\,13\,\mu$s region. The measured ionization spectrum for events passing all neutron-selection cuts and the drift time cut is shown in Figure~\ref{fig:dt} (bottom). The scaled random coincidence background, shown by the red points, is strictly composed of single electron events and entirely explains the peak at 1 detected electron in the $1.6\,$keV recoil spectrum. In the following analysis, we subtract this empirically-determined background from our spectra before analyzing the ionization yields.

The background-subtracted spectra at all energies are shown by the black data points in Figure~\ref{fig:spectra} at a drift field of 220$\,$V/cm. We see clear quantization of few-electron signals at the lowest energies, and are able to see the single-electron peaks even after backgrounds have been subtracted.

\subsection{Charge detection efficiencies}
\label{subsec:efficiencies}
There are two effects which reduce the measured ionization signal in the xenon TPC relative to the true ionization produced in the target: electron capture on impurities in the liquid, and incomplete extraction of electrons from the liquid into the gas electroluminescence region. 

The survival probability for electrons drifting in the liquid xenon is modeled by
\begin{equation}
    p_{_{sur}} = \exp{\left( -t_d/\tau \right)}
    \label{eq:purity}
\end{equation}
where $t_d$ is the drift time and $\tau$ is the electron lifetime in the liquid. 
We measure the electron lifetime in situ using the nuclear-recoil peak in BD channel 3. This channel is one of the larger LS detectors, and therefore has higher statistics that enable this calibration. We measure the peak position as a function of drift time. The peak position is binned in steps of 2~$\mu$s, and the resulting peaks are fitted to an exponential function to extract $\tau$. The measured lifetime using the full dataset is $\tau = 232^{+294}_{-82}$~$\mu$s. To explore time dependence, we also split the data into three time bins corresponding to different drift fields; the measured lifetime in each time bin was consistent with that given above (within the large uncertainties), and the relative changes among the three time bins were consistent with a constant electron lifetime correction over the course of the measurement campaign.

The electron extraction efficiency (EEE) is not measured directly in this work, but was measured previously in our detector and reported in Ref~\cite{LLNL_EEE2019}. The extraction probability for a given electron at our operating extraction field (6.24$\,$kV/cm) is taken from an empirical fit to the measurements. We obtain the value
$p_{_{extr}} \left(6.24\,\text{V/cm}\right) = 0.955^{+0.014}_{-0.017}$,
where the uncertainty is estimated by fitting to the data and their upper and lower $1$-$\sigma$ values. 

%% file: results_full_table.tex
{\setlength{\tabcolsep}{0.7em}
\begin{table*}[t]
    \def\arraystretch{1.7}
    \centering 
    \caption{Charge yield as a function of energy, measured at four different applied electric fields. The energies and their error bars are the mean and central 68-percentile of the simulated recoil energy distribution for single-scatter events in each BD channel. Uncertainties in the energy (denoted $\Delta E$) are divided into two parts: uncorrelated and correlated (explained in the text). The yields are given with the statistical uncertainties from the fit; systematic uncertainties in the yield are shown in separate columns/rows and are described in Section~\ref{subsec:systematics}.}
    \begin{tabular}{c|c|c c|c c c c|C{1.cm}|C{1.2cm}}
        \hline
        \hline
        BD          & $E$ (keV) & \multicolumn{2}{c|}{$\Delta E$} 
                    & \multicolumn{4}{c|}{$Q_y$} 
                    & \multirow[t]{2}{1.cm}{\centering Scaling syst.}
                    & \multirow[t]{2}{1.3cm}{\centering Modeling syst.} \\
                    
                    & & Uncorr. & Corr.
                    & 220V/cm
                    & 550V/cm
                    & 2.2kV/cm
                    & 6.3kV/cm & & \\
        \hline
        1           & $6.08\,^{+0.42}_{-0.52}$ 
                    & $\pm3.3$\% & $^{+0.04}_{-0.04}$ 
                    & $ 6.98\,^{+0.08}_{-0.08}$           
                    & $ 7.382\,^{+0.09}_{-0.11}$          
                    & $ 7.63\,^{+0.11}_{-0.14}$           
                    & $8.00\,^{+0.12}_{-0.13}$ 
                    & - & -\\
        2           & $4.65\,^{+0.25}_{-0.24}$ 
                    & $\pm0.8$\% & $^{+0.04}_{-0.04}$ 
                    & $6.99\,^{+0.12}_{-0.11}$            
                    & $ 7.46\,^{+0.14}_{-0.14}$           
                    & $ 7.46\,^{+0.14}_{-0.14}$           
                    & $7.95\,^{+0.21}_{-0.23}$ 
                    & - & -\\
        3           & $3.61\,^{+0.23}_{-0.22}$ 
                    & $\pm0.9$\% & $^{+0.04}_{-0.03}$ 
                    & $ 7.33\,^{+0.11}_{-0.13}$           
                    & $ 7.74\,^{+0.15}_{-0.15}$           
                    & $ 8.03\,^{+0.17}_{-0.15}$           
                    & $8.08\,^{+0.20}_{-0.20}$ 
                    & - & -\\
        4           & $2.95\,^{+0.21}_{-0.20}$ 
                    & $\pm1.1$\% & $^{+0.04}_{-0.03}$ 
                    & $ 6.96\,^{+0.10}_{-0.10}$           
                    & $ 7.53\,^{+0.13}_{-0.16}$           
                    & $ 7.77\,^{+0.11}_{-0.12}$           
                    & $8.17\,^{+0.11}_{-0.15}$ 
                    & - & -\\
        5           & $2.11\,^{+0.31}_{-0.28}$ 
                    & $\pm1.3$\% & $^{+0.03}_{-0.03}$ 
                    & $ 6.88\,^{+0.09}_{-0.09}$           
                    & $ 7.26\,^{+0.10}_{-0.10}$           
                    & $ 7.31\,^{+0.08}_{-0.10}$           
                    & $7.63\,^{+0.14}_{-0.09}$ 
                    & - & -\\
        6           & $1.61\,^{+0.16}_{-0.15}$ 
                    & $\pm1.5$\% & $^{+0.03}_{-0.03}$ 
                    & $ 6.89\,^{+0.21}_{-0.22}$           
                    & $ 7.13\,^{+0.16}_{-0.21}$           
                    & $ 7.36\,^{+0.15}_{-0.18}$           
                    & $7.764\,^{+0.18}_{-0.17}$ 
                    & - & -\\
        7           & $0.97\,^{+0.13}_{-0.11}$ 
                    & $\pm2.0$\% & $^{+0.03}_{-0.03}$ 
                    & $ 6.23\,^{+0.22}_{-0.18}$           
                    & $ 6.66\,^{+0.25}_{-0.32}$           
                    & $ 6.26\,^{+0.26}_{-0.21}$           
                    & $6.84\,^{+0.23}_{-0.29}$ 
                    & - & -\\
        8           & $0.93\,^{+0.12}_{-0.11}$ 
                    & $\pm2.0$\% & $^{+0.03}_{-0.03}$ 
                    & $ 6.32\,^{+0.23}_{-0.24}$           
                    & $ 6.48\,^{+0.27}_{-0.25}$           
                    & $ 6.47\,^{+0.26}_{-0.30}$           
                    & $6.84\,^{+0.27}_{-0.35}$ 
                    & - & -\\
        9           & $0.442\,^{+0.088}_{-0.074}$ 
                    & $\pm3.0$\% & $^{+0.016}_{-0.018}$ 
                    & $ 4.58\,^{+0.39}_{-0.38}$           
                    & $ 4.94\,^{+0.38}_{-0.36}$           
                    & $ 4.80\,^{+0.41}_{-0.43}$           
                    & $ 5.47\,^{+0.43}_{-0.43}$ 
                    & -5.9\% & 5.5\%\\
        10          & $0.296\,^{+0.074}_{-0.062}$ 
                    & $\pm3.6$\% & $^{+0.018}_{-0.014}$ 
                    & $ 3.47\,^{+0.41}_{-0.40}$           
                    & $ 4.50\,^{+0.48}_{-0.45}$           
                    & $ 4.31\,^{+0.40}_{-0.37}$           
                    & $ 4.46\,^{+0.50}_{-0.50}$ 
                    & +6.4\% & 11.0\%\\
        \hline
        \multicolumn{3}{l}{Electron lifetime systematic} & &
        $\pm2.9$\% & $\pm2.5$\% & $\pm2.1$\% & - \\
        \multicolumn{3}{l}{Extraction efficiency systematic} & &
        \multicolumn{4}{c|}{$+2.0\%$ / $-1.5\%$} \\
        \hline
        \hline
            
    \end{tabular}
    \label{tab:yields}
\end{table*}
}

Our strategy for measuring the ionization yield from our data consists of fully simulating the ionization spectra under different assumed yields and then comparing to data. Events are grouped by the applied field and BD channel into 40 spectra, which we fit independently to measure both the energy- and field-dependence. Fitting is accomplished using a Bayesian analysis, in which the posterior distribution is sampled directly using a Markov Chain Monte Carlo (MCMC) technique. We describe the simulation technique, fitting algorithm, and results below. 

\subsection{Simulation of ionization spectra}
We generate simulated recoil energy spectra using a Geant4-based Monte Carlo model of the experimental setup, which includes detailed geometries of the collimator, the xenon detector and its vacuum chamber, and the liquid scintillator tagging detectors. Our simulation is built using the BACCARAT application, a detector-independent branch of LUXSim~\cite{LUXSim_Paper}. Primary neutrons are generated from a point source at the location of the LiF target, and are emitted in a spherically-uniform cone of opening angle 7~degrees (neutrons with larger opening angles make up less than 2\% of the observed events after selection cuts, and are considered negligible). Neutron energies are drawn from a normal distribution with a mean energy of 579~keV and a 1-$\sigma$ width of 10~keV. Energy deposits are recorded in all materials included in the simulation, and events are tagged as either single-scatter (if they scatter only once in the xenon TPC before interacting in a liquid scintillator detector) or multiple-scatter (if they scatter in any passive material before reaching the liquid scintillator). We apply TOF- and $x/y$-position cuts to the simulation to mimic our analysis of real data; we only accept events within the FWHM of the neutron TOF peak, and which have energy deposited within 1.3$\,$cm of the center of the xenon TPC. The energy spectra of the remaining events include both single- and multiple-scatter events, and represent the distribution of recoil energies expected in our measurements. These events undergo further event-by-event fluctuations to model the ionization process and detection efficiencies, described below.

The ionization process is more complicated than a simple Poisson process: competing effects from Fano statistics and the fluctuations in electron/ion recombination may serve to narrow or broaden the distribution, respectively~\cite{DokeFanoFactors,LUX2017_Yield,LUXTritium}. Therefore, we model ionization at $\sim$1$\,$keV and above by drawing the number of ionized electrons ($N_e$) from a normal distribution with a mean equal to the yield times the energy:
\begin{equation}
    \mu_{_{N_e}} = Q_y \times E
    \label{eq:mean_electrons}
\end{equation}
and a width equal to:
\begin{equation}
    \sigma_{_{N_e}} = \omega \sqrt{Q_y \times E}
    \label{eq:sig_electrons}
\end{equation}
where $Q_y$ is the absolute ionization yield (electrons per keV). The width parameter $\omega$ is an empirical variable that quantifies the width of the distribution relative to a Poisson process.
The sampled value is then rounded to the nearest integer to model discrete electron counting. 
In the two lowest-energy BD channels, we expect a significant fraction of the events to produce zero ionization electrons, so the distribution is truncated and may be skewed. A gaussian model is not expected to correctly capture these features. We instead model ionization at these energies as a simple Poisson process, with a mean number of ionized electrons given by Equation~\ref{eq:mean_electrons}. This simplifying assumption introduces a systematic uncertainty which we discuss in Section~\ref{subsec:systematics}.

We account for the electron lifetime and electron extraction losses by resampling the number of ionized electrons from successive binomial distributions. We first calculate the drift time for each simulated event by dividing the event depth in $z$ by the expected drift velocity at the measured drift field. Drift velocity as a function of field is obtained by a fit to the measurements in Ref.~\cite{EXO2017_EDrift}. The survival probability is then given by Eq.~\ref{eq:purity}. The probability of electron extraction is given by the fit described in Section~\ref{subsec:efficiencies}. After resampling the total number of electrons from each of these binomial distributions, we are left with a simulated number of detected electrons.

Finally, the simulated signals are smeared by the measured single-electron resolution: a normal distribution with $\sigma=20.5\%$. After all of the above fluctuations have been applied to each event, the resulting spectra can be directly compared to our data.

\subsection{Fits}

To fit the simulated distributions to the data, we vary the ionization yield $Q_y$ and the width parameter $\omega$ described above. We add an additional normalization factor $A$ to the fit, which parameterizes the total number of events in our simulated distributions relative to the data. Thus there are three free parameters which must be estimated simultaneously to produce a measurement of the ionization yield at each energy.

We construct a likelihood function which takes the number of events in each data bin as a Poisson-distributed random variable with a mean equal to the number of events in the simulation bins, scaled by the normalization factor $A$. We perform a Bayesian analysis, where the likelihood is related to the posterior distribution of the parameters by Bayes' theorem,
\begin{equation}
L(A,Q_y,\omega\,|\,data) \propto P(data\,|\,A,Q_y,\omega) \times P(A,Q_y,\omega)
\end{equation}
where we have assumed the probability of measuring a specific set of data ($P(data)$) is uniform and can be neglected. The probability distribution of the values of the parameters $P(A,Q_y,\omega)$ is initially taken as uniform as well, but can be modified to include priors on the parameter values. This is utilized in our analysis of the two lowest-energy BD channels as described below. 

The posterior probability distribution is sampled with a Markov Chain Monte Carlo (MCMC) sampler using the Metropolis-Hastings algorithm. 
At each step, the simulation events are fully reprocessed through the ionization and efficiencies framework to account for correlated changes to the spectra induced by new values of the parameters. The MCMC sampler is run for 5000 steps, and we discard the first 500 samples to avoid including highly correlated samples which are generated as the algorithm initially converges on the highest-likelihood region of the parameter space. The central values and the uncertainties in each parameter are computed by marginalizing the resulting sample distributions. The median value of the sample distribution in each parameter is reported as the ``best-fit", and the central 68th-percentile is reported as its confidence interval.

Fitting proceeds in two stages. We first fit each of the spectra for BD 1--8, allowing $A$, $Q_y$, and $\omega$ to float unconstrained in the fit. We then find the mean and standard deviation of the values of $A$ for all of the fits. This is used to apply a gaussian prior on $A$ for the two lowest-energy bins. This constraint allows us to fit a distribution to our data which includes the possibility of 0-ionization-electron events (which are below our trigger threshold), allowing us to report an average yield at these energies.

The multiple-scatter background has a broad energy distribution, and the charge yield is known to vary with energy. We therefore perform our fits iteratively. In the first stage, the yield is assumed to be constant in energy for each spectrum, and the best-fit yield is extracted for each channel. The single-scatter peak energy for each channel is used to construct a preliminary measurement of the yield-vs-energy, which is fit to an empirical function to interpolate between different energies. In the second stage, the empirical yield-vs-energy relationship is used to calculate the ionization for multiple-scatter events, while the yield for single-scatter events (which are narrowly peaked in energy) remains a single free parameter and floats in the MCMC sampling procedure.

\subsection{Results}

\begin{figure}[t!]
    \centering
        \includegraphics[width=0.448\textwidth]{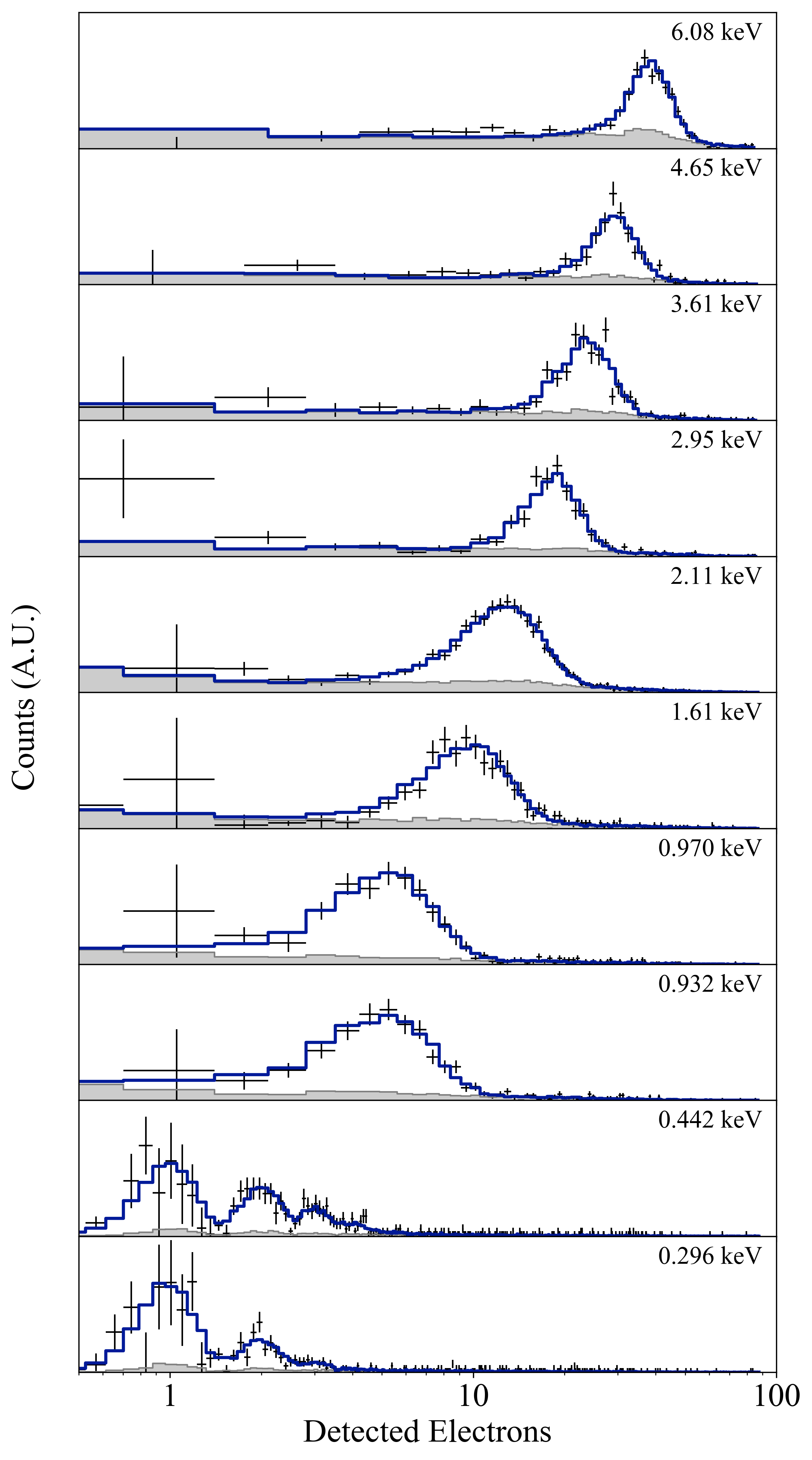}
        \caption{Measured ionization spectra at a drift field of 200~V/cm (black histogram) overlaid on the best-fit spectra from simulations. The simulated distributions show both the multiple-scattering background (grey shaded) and the full simulated distribution with single- and multiple-scatters summed together (blue curve).}
        \label{fig:spectra}
\end{figure}

\begin{figure}[t]
    \centering
        \includegraphics[width=0.45\textwidth]{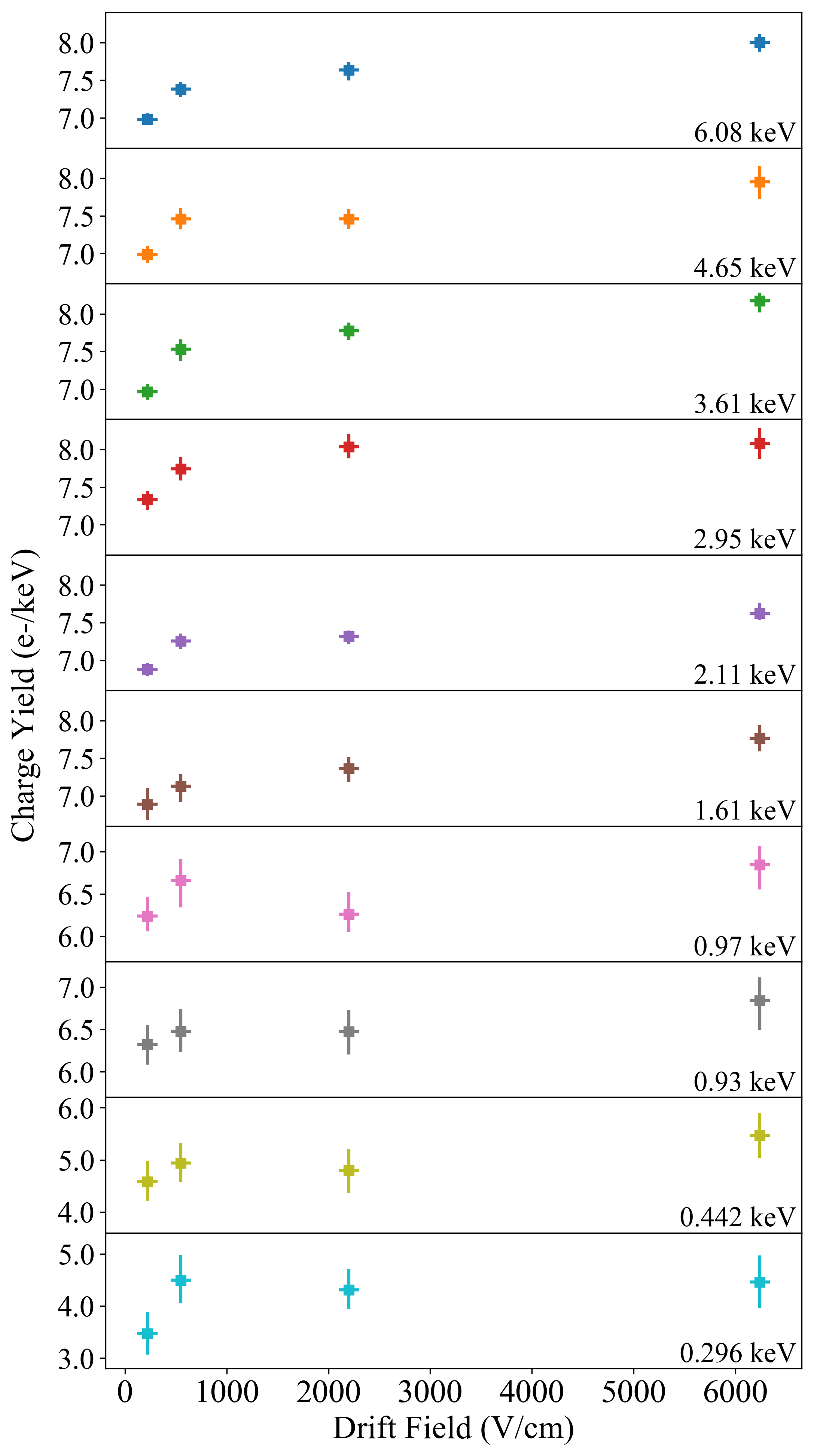}
        \caption{Field-dependence of the measured charge yield at all energies. We observe a significant dependence on the electric field, in contrast with recent results from Ref.~\cite{Aprile2018_Yields}.}
        \label{fig:field_dependence}
\end{figure}

Results from the fit are tabulated in Table~\ref{tab:yields}. The simulated distributions at each recoil energy, with the best-fit values for $Q_y$, $\omega$, and $A$ at a drift field of $\sim$220~V/cm, are plotted against against our data in Figure~\ref{fig:spectra}. 
We also show the ionization yield as a function of drift field for each energy bin in Figure~\ref{fig:field_dependence}. We find a significant dependence on electric field for all measurements above 1$\,$keV, with the charge yield increasing by 10--15\% from 220$\,$V/cm to 6300$\,$V/cm. At energies below 1$\,$keV the statistical uncertainties are too large to conclusively determine if this field-dependence continues. Our observations suggest that recombination does indeed play a role in the measured signal strength at low energies, and that an increasing drift field may provide a small increase the sensitivity of detectors searching for low energy nuclear recoils.

The goodness-of-fit is quantified using a likelihood ratio test statistic, where
\begin{equation}
    \chi^2 \approx -2\ln{ \lambda }
\end{equation}
where $\lambda$ is the ratio of two likelihoods for two different models. As our alternative hypothesis we use the so-called ``saturated model", in which the likelihood is calculated assuming that the expected value in each bin is equal to the observed value~\cite{BakerCousins1983_SaturatedModels}. We assume that the number of counts in each channel is sufficient that this can be approximated as a $\chi^2$-distributed variable. 
The resulting p-values from a standard $\chi^2$ goodness-of-fit test approximately give the probability that a dataset generated by our model would produce a likelihood at least as extreme as our observation when compared to our model.
We find that $\sim$90\% of the fits are consistent with our data at a level of $p>0.05$, with no discernible pattern or systematic deviation in the remaining fits. All of the 40 fits are consistent with our data at a level of $p>0.01$. We therefore conclude that our simulation and charge yield model is a good description of our data.


\subsection{Systematic uncertainties}
\label{subsec:systematics}

Uncertainty in the positions of each of the various detectors contributes to an uncertainty in the recoil energy in each channel. Due to the energy dependence in Eqs.~\ref{eq:mean_electrons} and \ref{eq:sig_electrons}, this in turn becomes an uncertainty in the measured ionization yield. These uncertainties can be subdivided into two categories: uncorrelated and correlated. The first is due to uncertainties in the positions of each of the LS detectors, which we assume are independent in each channel. These are listed in Table~\ref{tab:yields} as ``$\Delta E$ uncorr.". The second is due to the uncertainty in the position of the xenon detector; if the TPC is shifted slightly to the right of the incoming beam, the angles in channels to the right (left) will be slightly smaller (larger), meaning that we will have overestimated (underestimated) the average recoil energy in that channel. Because this affects all channels simultaneously, we list this under ``$\Delta E$ corr." in Table~\ref{tab:yields}.

Our analysis is also subject to two independent overall scaling uncertainties, which stem from our assumptions of the electron lifetime and the electron extraction efficiency. The former is measured in-situ, as described in Section~\ref{subsec:calibration}. To evaluate its effect on our measurement of $Q_y$, we regenerate our simulated spectra using the electron lifetime at its $\pm1\,\sigma$ values and repeat the fitting procedure. We find deviations of $<3\%$; the exact value for each drift field dataset is listed in Table~\ref{tab:yields}. For events above the extraction grid, the drift time is so short that this effect is negligible. The uncertainty in the electron extraction efficiency is propagated into our $Q_y$ measurement in the same way: we take $p_{_{extr}}$ at its $\pm1\,\sigma$ extremes and re-fit the spectra. We find an overall scaling uncertainty of O(2\%), which applies to all of our measured $Q_y$ values. The uncertainty from variations in the single-electron response of the xenon TPC, mentioned in Section~\ref{subsec:calibration}, produces an additional scaling uncertainty of 0.5\%. This is significantly smaller than those mentioned above and is therefore neglected in our final error analysis.

Finally, there are two systematic uncertainties which affect only the lowest energy recoils. The first is due to our assumption that the mean scaling factor $A$ fitted at higher energies can be used to constrain $A$ for the two lowest energy spectra. This is equivalent to assuming that our simulation correctly models a) the angular dependence of the neutron elastic scattering cross section on xenon and b) the exact position of all shielding components and detector positions relative to the beam. We address a) by fitting a line with variable slope to the fitted values of $A$ as a function of energy for all spectra above 0.5~keV. This line is then extrapolated to the two lower energy bins. This models a systematic linear deviation from the angular scattering cross section. Refitting the spectra at 0.4 and 0.3~keV under these assumptions leads to a $3.2\%$ and $3.4\%$ increase in the measured yield, respectively, across all drift fields. We address b) by splitting the data into channels to the left and right of the incoming beam, and finding an average $A$ for each side independently. This models any left/right dependence in our experiment from a misalignment of the shielding or detectors. We find that the best-fit $A$ on each side deviates from the global average by approximate $1$-$\sigma$ in either direction, and leads to a $-5.9\%$ and $+6.4\%$ change in the charge yields at 0.4 and 0.3~keV, respectively (the lowest energy detectors are on opposite sides of the beam). As the latter effect is larger, we report this as the scaling systematic uncertainty in Table~\ref{tab:yields}. 

The second systematic in the two lowest energies is due to our choice of using a Poisson distribution to model charge production. We do not have a physical model for this distribution; energy conservation in the ionization process and fluctuations in electron/ion recombination lead to competing effects which serve to narrow or broaden the distribution with respect to a Poisson process, respectively. A first-principles derivation of these effects is beyond the scope of this work. We can, however, estimate our uncertainty due to the choice of model by evaluating the ionization yield from data directly. Using the simulations and the average best-fit values for the normalization parameter $A$ at higher energies ($>1\,$keV), we predict the number of events that will be present in the two lowest energy spectra. We then use this to estimate the number of events that would produce zero ionization electrons. This allows us to calculate the average of the measured ionization signal in a model-independent fashion, and compare with our Poisson fits. We find that, in all cases, the calculated yield is in agreement with the average yield within our statistical uncertainties. We take the maximum deviation as a systematic uncertainty due to our use of the Poisson model in our fits. This is shown as a systematic uncertainty in the rightmost column of Table~\ref{tab:yields}.

%% file: discussion.tex
\begin{figure}[t!]
    \centering
        \includegraphics[width=0.45\textwidth]{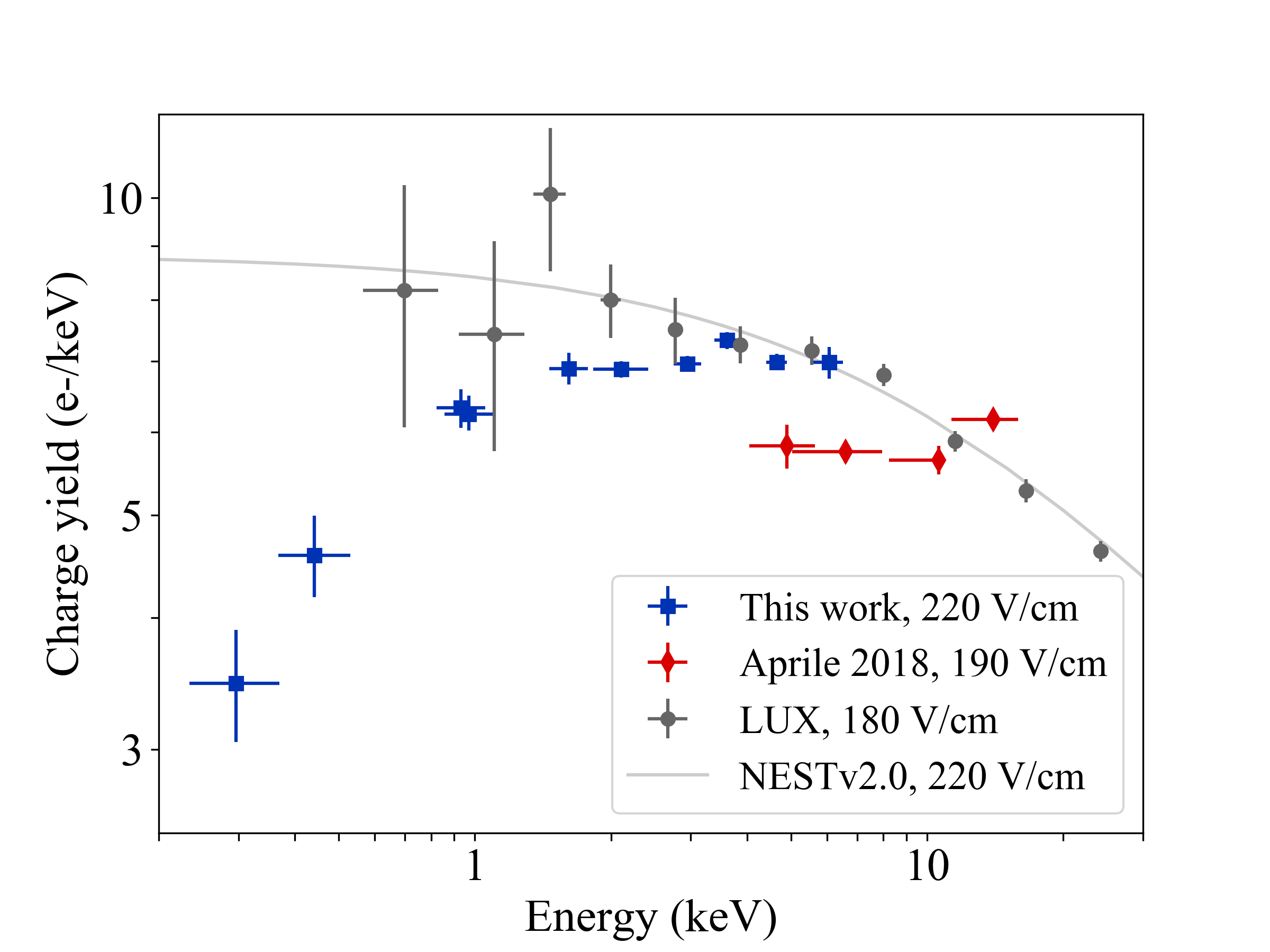}
        \caption{Measurements of the energy-dependence of ionization yield in liquid xenon at 220$\,$V/cm, compared with recent measurements from the LUX Collaboration~\cite{LUXDD} and Aprile~\emph{et al.}~\cite{Aprile2018_Yields} made at similar drift fields. We also show the ionization yield currently implemented in the Noble Element Simulation Technique software package (version 2.0)~\cite{NESTZenodo}}
        \label{fig:energy_dependence}
\end{figure}

We compare the energy dependence of $Q_y$ directly to two other recent measurements in Figure~\ref{fig:energy_dependence}. Our measurements are in agreement with the LUX data at a similar electric field~\cite{LUXDD}, but have smaller uncertainties and extend the calibration to lower energies. We find that the ionization yield decreases significantly below 1$\,$keV. 
We also compare our results to the Noble Element Simulation Technique (NEST), a publicly-available software package that has become the standard benchmark for modeling ionization and scintillation production in liquid xenon~\cite{NESTpaper,NESTpaper2}. The ionization models in NEST are tuned to best describe available data, and the most recent nuclear recoil model is therefore primarily constrained by the LUX measurements in the $\sim$1$\,$keV regime~\cite{NESTZenodo}. When extrapolated to lower energies, this model overpredicts the ionization production. Our results can inform the next generation of the NEST simulation code, and will help improve modeling of ultra-low-energy ionization processes in liquid xenon throughout the larger community.

We also find a non-negligible dependence of the charge yield on the applied electric field, in contrast with the conclusion of Ref.~\cite{Aprile2018_Yields}, but consistent with results above 25~keV~\cite{ColumbiaQy2006}. The measured field dependence is weaker than that implemented in the most recent version of NEST, and will again help inform future modeling efforts.

At the lowest energy in our measurement, we find a yield of 3.5--4.5$\,$e$^{-}$/keV, which corresponds to an average signal strength of 1.1--1.4 ionized electrons. This is the smallest nuclear recoil signal measured to date, and approaches the fundamental limit of nuclear recoil detection in liquid xenon. Detectors that hope to be sensitive to recoils at energies lower than 300$\,$eV must have thresholds of 1--2 detected electrons and rely on upward fluctuations in the ionization signals. The average recoil energy for 5$\,$GeV WIMPs, solar $^8$B neutrinos, and reactor antineutrinos in a liquid xenon detector are expected to be 331$\,$eV, 433$\,$eV, and 67$\,$eV, with endpoints at 2.4$\,$keV, 4$\,$keV, and 1.6$\,$keV, respectively. Our measurements provide crucial data that can be used to accurately calculate the sensitivities of detectors for applications in low-mass WIMP searches and searches for neutrino signals via the CE$\nu$NS interaction.

%% file: conclusion.tex
This work describes a new measurement of the the nuclear recoil ionization yield in liquid xenon between 0.3 and 6.1$\,keV$ using fixed-angle neutron scattering. With strong background rejection techniques, high electron extraction efficiency, and single-extracted-electron triggering capabilities, we are able for the first time to measure nuclear recoils which produce only a single ionization electron. Our measurements improve the precision of low energy nuclear recoil calibrations in the 1$\,$keV regime, and extend the calibration to lower energies than have previously been probed. We also measure this property as a function of applied electric field, extending our work's applicability to a wide range of experimental operating parameters. This provides a new detector-independent calibration of liquid xenon ionization production in a energy regime that is critical for many of the scientific applications for dual-phase xenon TPCs.

%% file: XeRecoil-prc.bbl
\begin{thebibliography}{33}
\expandafter\ifx\csname natexlab\endcsname\relax\def\natexlab#1{#1}\fi
\expandafter\ifx\csname bibnamefont\endcsname\relax
  \def\bibnamefont#1{#1}\fi
\expandafter\ifx\csname bibfnamefont\endcsname\relax
  \def\bibfnamefont#1{#1}\fi
\expandafter\ifx\csname citenamefont\endcsname\relax
  \def\citenamefont#1{#1}\fi
\expandafter\ifx\csname url\endcsname\relax
  \def\url#1{\texttt{#1}}\fi
\expandafter\ifx\csname urlprefix\endcsname\relax\def\urlprefix{URL }\fi
\providecommand{\bibinfo}[2]{#2}
\providecommand{\eprint}[2][]{\url{#2}}

\bibitem[{\citenamefont{Akerib et~al.}(2017{\natexlab{a}})\citenamefont{Akerib,
  Alsum, Ara\'ujo, Bai, Bailey, Balajthy, Beltrame, Bernard, Bernstein,
  Biesiadzinski et~al.}}]{LUXRun4PRL}
\bibinfo{author}{\bibfnamefont{D.~S.} \bibnamefont{Akerib}},
  \bibinfo{author}{\bibfnamefont{S.}~\bibnamefont{Alsum}},
  \bibinfo{author}{\bibfnamefont{H.~M.} \bibnamefont{Ara\'ujo}},
  \bibinfo{author}{\bibfnamefont{X.}~\bibnamefont{Bai}},
  \bibinfo{author}{\bibfnamefont{A.~J.} \bibnamefont{Bailey}},
  \bibinfo{author}{\bibfnamefont{J.}~\bibnamefont{Balajthy}},
  \bibinfo{author}{\bibfnamefont{P.}~\bibnamefont{Beltrame}},
  \bibinfo{author}{\bibfnamefont{E.~P.} \bibnamefont{Bernard}},
  \bibinfo{author}{\bibfnamefont{A.}~\bibnamefont{Bernstein}},
  \bibinfo{author}{\bibfnamefont{T.~P.} \bibnamefont{Biesiadzinski}},
  \bibnamefont{et~al.} (\bibinfo{collaboration}{LUX Collaboration}),
  \href{http://dx.doi.org/10.1103/PhysRevLett.118.021303}{\bibinfo{journal}{Phys.
  Rev. Lett.}, \textbf{\bibinfo{volume}{118}},
  \bibinfo{pages}{021303}\bibinfo{year}{
  (\bibinfo{year}{2017}{\natexlab{a}})}}.

\bibitem[{\citenamefont{Aprile et~al.}(2018{\natexlab{a}})\citenamefont{Aprile,
  Aalbers, Agostini, Alfonsi, Althueser, Amaro, Anthony, Arneodo, Baudis,
  Bauermeister et~al.}}]{XENON1T_2018}
\bibinfo{author}{\bibfnamefont{E.}~\bibnamefont{Aprile}},
  \bibinfo{author}{\bibfnamefont{J.}~\bibnamefont{Aalbers}},
  \bibinfo{author}{\bibfnamefont{F.}~\bibnamefont{Agostini}},
  \bibinfo{author}{\bibfnamefont{M.}~\bibnamefont{Alfonsi}},
  \bibinfo{author}{\bibfnamefont{L.}~\bibnamefont{Althueser}},
  \bibinfo{author}{\bibfnamefont{F.~D.} \bibnamefont{Amaro}},
  \bibinfo{author}{\bibfnamefont{M.}~\bibnamefont{Anthony}},
  \bibinfo{author}{\bibfnamefont{F.}~\bibnamefont{Arneodo}},
  \bibinfo{author}{\bibfnamefont{L.}~\bibnamefont{Baudis}},
  \bibinfo{author}{\bibfnamefont{B.}~\bibnamefont{Bauermeister}},
  \bibnamefont{et~al.} (\bibinfo{collaboration}{XENON Collaboration}),
  \href{http://dx.doi.org/10.1103/PhysRevLett.121.111302}{\bibinfo{journal}{Phys.
  Rev. Lett.}, \textbf{\bibinfo{volume}{121}},
  \bibinfo{pages}{111302}\bibinfo{year}{
  (\bibinfo{year}{2018}{\natexlab{a}})}}.

\bibitem[{\citenamefont{Tan et~al.}(2016)\citenamefont{Tan, Xiao, Cui, Chen,
  Chen, Fang, Fu, Giboni, Giuliani, Gong et~al.}}]{PandaXSpinIndependent2016}
\bibinfo{author}{\bibfnamefont{A.}~\bibnamefont{Tan}},
  \bibinfo{author}{\bibfnamefont{M.}~\bibnamefont{Xiao}},
  \bibinfo{author}{\bibfnamefont{X.}~\bibnamefont{Cui}},
  \bibinfo{author}{\bibfnamefont{X.}~\bibnamefont{Chen}},
  \bibinfo{author}{\bibfnamefont{Y.}~\bibnamefont{Chen}},
  \bibinfo{author}{\bibfnamefont{D.}~\bibnamefont{Fang}},
  \bibinfo{author}{\bibfnamefont{C.}~\bibnamefont{Fu}},
  \bibinfo{author}{\bibfnamefont{K.}~\bibnamefont{Giboni}},
  \bibinfo{author}{\bibfnamefont{F.}~\bibnamefont{Giuliani}},
  \bibinfo{author}{\bibfnamefont{H.}~\bibnamefont{Gong}}, \bibnamefont{et~al.}
  (\bibinfo{collaboration}{PandaX-II Collaboration}),
  \href{http://dx.doi.org/10.1103/PhysRevLett.117.121303}{\bibinfo{journal}{Phys.
  Rev. Lett.}, \textbf{\bibinfo{volume}{117}},
  \bibinfo{pages}{121303}\bibinfo{year}{ (\bibinfo{year}{2016})}}.

\bibitem[{\citenamefont{Akerib et~al.}(2016{\natexlab{a}})\citenamefont{Akerib,
  Ara\'ujo, Bai, Bailey, Balajthy, Beltrame, Bernard, Bernstein, Biesiadzinski,
  Boulton et~al.}}]{LUXRun3Reanalysis}
\bibinfo{author}{\bibfnamefont{D.~S.} \bibnamefont{Akerib}},
  \bibinfo{author}{\bibfnamefont{H.~M.} \bibnamefont{Ara\'ujo}},
  \bibinfo{author}{\bibfnamefont{X.}~\bibnamefont{Bai}},
  \bibinfo{author}{\bibfnamefont{A.~J.} \bibnamefont{Bailey}},
  \bibinfo{author}{\bibfnamefont{J.}~\bibnamefont{Balajthy}},
  \bibinfo{author}{\bibfnamefont{P.}~\bibnamefont{Beltrame}},
  \bibinfo{author}{\bibfnamefont{E.~P.} \bibnamefont{Bernard}},
  \bibinfo{author}{\bibfnamefont{A.}~\bibnamefont{Bernstein}},
  \bibinfo{author}{\bibfnamefont{T.~P.} \bibnamefont{Biesiadzinski}},
  \bibinfo{author}{\bibfnamefont{E.~M.} \bibnamefont{Boulton}},
  \bibnamefont{et~al.} (\bibinfo{collaboration}{LUX Collaboration}),
  \href{http://dx.doi.org/10.1103/PhysRevLett.116.161301}{\bibinfo{journal}{Phys.
  Rev. Lett.}, \textbf{\bibinfo{volume}{116}},
  \bibinfo{pages}{161301}\bibinfo{year}{
  (\bibinfo{year}{2016}{\natexlab{a}})}}.

\bibitem[{\citenamefont{Angle et~al.}(2011)\citenamefont{Angle, Aprile,
  Arneodo, Baudis, Bernstein, Bolozdynya, Coelho, Dahl, DeViveiros, Ferella
  et~al.}}]{XENON10_S2only}
\bibinfo{author}{\bibfnamefont{J.}~\bibnamefont{Angle}},
  \bibinfo{author}{\bibfnamefont{E.}~\bibnamefont{Aprile}},
  \bibinfo{author}{\bibfnamefont{F.}~\bibnamefont{Arneodo}},
  \bibinfo{author}{\bibfnamefont{L.}~\bibnamefont{Baudis}},
  \bibinfo{author}{\bibfnamefont{A.}~\bibnamefont{Bernstein}},
  \bibinfo{author}{\bibfnamefont{A.~I.} \bibnamefont{Bolozdynya}},
  \bibinfo{author}{\bibfnamefont{L.~C.~C.} \bibnamefont{Coelho}},
  \bibinfo{author}{\bibfnamefont{C.~E.} \bibnamefont{Dahl}},
  \bibinfo{author}{\bibfnamefont{L.}~\bibnamefont{DeViveiros}},
  \bibinfo{author}{\bibfnamefont{A.~D.} \bibnamefont{Ferella}},
  \bibnamefont{et~al.} (\bibinfo{collaboration}{XENON10 Collaboration}),
  \href{http://dx.doi.org/10.1103/PhysRevLett.107.051301}{\bibinfo{journal}{Phys.
  Rev. Lett.}, \textbf{\bibinfo{volume}{107}},
  \bibinfo{pages}{051301}\bibinfo{year}{ (\bibinfo{year}{2011})}}.

\bibitem[{\citenamefont{Akimov et~al.}(2017)\citenamefont{Akimov, Albert, An,
  Awe, Barbeau, Becker, Belov, Brown, Bolozdynya, Cabrera-Palmer
  et~al.}}]{COHERENT2017_CENNS}
\bibinfo{author}{\bibfnamefont{D.}~\bibnamefont{Akimov}},
  \bibinfo{author}{\bibfnamefont{J.~B.} \bibnamefont{Albert}},
  \bibinfo{author}{\bibfnamefont{P.}~\bibnamefont{An}},
  \bibinfo{author}{\bibfnamefont{C.}~\bibnamefont{Awe}},
  \bibinfo{author}{\bibfnamefont{P.~S.} \bibnamefont{Barbeau}},
  \bibinfo{author}{\bibfnamefont{B.}~\bibnamefont{Becker}},
  \bibinfo{author}{\bibfnamefont{V.}~\bibnamefont{Belov}},
  \bibinfo{author}{\bibfnamefont{A.}~\bibnamefont{Brown}},
  \bibinfo{author}{\bibfnamefont{A.}~\bibnamefont{Bolozdynya}},
  \bibinfo{author}{\bibfnamefont{B.}~\bibnamefont{Cabrera-Palmer}},
  \bibnamefont{et~al.},
  \href{http://dx.doi.org/10.1126/science.aao0990}{\bibinfo{journal}{Science},
  \bibinfo{year}{ (\bibinfo{year}{2017})}}, ISSN \bibinfo{issn}{0036-8075}.

\bibitem[{\citenamefont{Dutta et~al.}(2016{\natexlab{a}})\citenamefont{Dutta,
  Mahapatra, Strigari, and Walker}}]{DuttaZprime2016}
\bibinfo{author}{\bibfnamefont{B.}~\bibnamefont{Dutta}},
  \bibinfo{author}{\bibfnamefont{R.}~\bibnamefont{Mahapatra}},
  \bibinfo{author}{\bibfnamefont{L.~E.} \bibnamefont{Strigari}},
  \bibnamefont{and} \bibinfo{author}{\bibfnamefont{J.~W.}
  \bibnamefont{Walker}},
  \href{http://dx.doi.org/10.1103/PhysRevD.93.013015}{\bibinfo{journal}{Phys.
  Rev. D}, \textbf{\bibinfo{volume}{93}},
  \bibinfo{pages}{013015}\bibinfo{year}{
  (\bibinfo{year}{2016}{\natexlab{a}})}}.

\bibitem[{\citenamefont{Dutta et~al.}(2016{\natexlab{b}})\citenamefont{Dutta,
  Gao, Kubik, Mahapatra, Mirabolfathi, Strigari, and
  Walker}}]{DuttaSterile2016}
\bibinfo{author}{\bibfnamefont{B.}~\bibnamefont{Dutta}},
  \bibinfo{author}{\bibfnamefont{Y.}~\bibnamefont{Gao}},
  \bibinfo{author}{\bibfnamefont{A.}~\bibnamefont{Kubik}},
  \bibinfo{author}{\bibfnamefont{R.}~\bibnamefont{Mahapatra}},
  \bibinfo{author}{\bibfnamefont{N.}~\bibnamefont{Mirabolfathi}},
  \bibinfo{author}{\bibfnamefont{L.~E.} \bibnamefont{Strigari}},
  \bibnamefont{and} \bibinfo{author}{\bibfnamefont{J.~W.}
  \bibnamefont{Walker}},
  \href{http://dx.doi.org/10.1103/PhysRevD.94.093002}{\bibinfo{journal}{Phys.
  Rev. D}, \textbf{\bibinfo{volume}{94}},
  \bibinfo{pages}{093002}\bibinfo{year}{
  (\bibinfo{year}{2016}{\natexlab{b}})}}.

\bibitem[{\citenamefont{Anderson et~al.}(2012)\citenamefont{Anderson, Conrad,
  Figueroa-Feliciano, Ignarra, Karagiorgi, Scholberg, Shaevitz, and
  Spitz}}]{Anderson2012}
\bibinfo{author}{\bibfnamefont{A.~J.} \bibnamefont{Anderson}},
  \bibinfo{author}{\bibfnamefont{J.~M.} \bibnamefont{Conrad}},
  \bibinfo{author}{\bibfnamefont{E.}~\bibnamefont{Figueroa-Feliciano}},
  \bibinfo{author}{\bibfnamefont{C.}~\bibnamefont{Ignarra}},
  \bibinfo{author}{\bibfnamefont{G.}~\bibnamefont{Karagiorgi}},
  \bibinfo{author}{\bibfnamefont{K.}~\bibnamefont{Scholberg}},
  \bibinfo{author}{\bibfnamefont{M.~H.} \bibnamefont{Shaevitz}},
  \bibnamefont{and} \bibinfo{author}{\bibfnamefont{J.}~\bibnamefont{Spitz}},
  \href{http://dx.doi.org/10.1103/PhysRevD.86.013004}{\bibinfo{journal}{Phys.
  Rev. D}, \textbf{\bibinfo{volume}{86}},
  \bibinfo{pages}{013004}\bibinfo{year}{ (\bibinfo{year}{2012})}}.

\bibitem[{\citenamefont{Cadeddu et~al.}(2018)\citenamefont{Cadeddu, Giunti,
  Kouzakov, Li, Studenikin, and Zhang}}]{COHERENT_NuChargeRadii}
\bibinfo{author}{\bibfnamefont{M.}~\bibnamefont{Cadeddu}},
  \bibinfo{author}{\bibfnamefont{C.}~\bibnamefont{Giunti}},
  \bibinfo{author}{\bibfnamefont{K.~A.} \bibnamefont{Kouzakov}},
  \bibinfo{author}{\bibfnamefont{Y.~F.} \bibnamefont{Li}},
  \bibinfo{author}{\bibfnamefont{A.~I.} \bibnamefont{Studenikin}},
  \bibnamefont{and} \bibinfo{author}{\bibfnamefont{Y.~Y.} \bibnamefont{Zhang}},
   \href{http://dx.doi.org/10.1103/PhysRevD.98.113010}{\bibinfo{journal}{Phys.
  Rev. D}, \textbf{\bibinfo{volume}{98}},
  \bibinfo{pages}{113010}\bibinfo{year}{ (\bibinfo{year}{2018})}}.

\bibitem[{\citenamefont{Cañas et~al.}(2018)\citenamefont{Cañas, Garcés,
  Miranda, and Parada}}]{Canas_CoherentWeakMixingAngle2018}
\bibinfo{author}{\bibfnamefont{B.}~\bibnamefont{Cañas}},
  \bibinfo{author}{\bibfnamefont{E.}~\bibnamefont{Garcés}},
  \bibinfo{author}{\bibfnamefont{O.}~\bibnamefont{Miranda}}, \bibnamefont{and}
  \bibinfo{author}{\bibfnamefont{A.}~\bibnamefont{Parada}},
  \href{http://dx.doi.org/https://doi.org/10.1016/j.physletb.2018.07.049}{\bibinfo{journal}{Physics
  Letters B}, \textbf{\bibinfo{volume}{784}}, \bibinfo{pages}{159
  }\bibinfo{year}{ (\bibinfo{year}{2018})}}, ISSN \bibinfo{issn}{0370-2693}.

\bibitem[{\citenamefont{Akerib et~al.}(2018)}]{LZSensitivityPaper}
\bibinfo{author}{\bibfnamefont{D.~S.} \bibnamefont{Akerib}}
  \bibnamefont{et~al.} (\bibinfo{collaboration}{LUX-ZEPLIN}),
  \href{http://arxiv.org/abs/1802.06039}{:1802.06039\bibinfo{year}{
  (\bibinfo{year}{2018})}}.

\bibitem[{\citenamefont{Lang et~al.}(2016)\citenamefont{Lang, McCabe, Reichard,
  Selvi, and Tamborra}}]{Lang2016_SupernovaXe}
\bibinfo{author}{\bibfnamefont{R.~F.} \bibnamefont{Lang}},
  \bibinfo{author}{\bibfnamefont{C.}~\bibnamefont{McCabe}},
  \bibinfo{author}{\bibfnamefont{S.}~\bibnamefont{Reichard}},
  \bibinfo{author}{\bibfnamefont{M.}~\bibnamefont{Selvi}}, \bibnamefont{and}
  \bibinfo{author}{\bibfnamefont{I.}~\bibnamefont{Tamborra}},
  \href{http://dx.doi.org/10.1103/PhysRevD.94.103009}{\bibinfo{journal}{Phys.
  Rev. D}, \textbf{\bibinfo{volume}{94}},
  \bibinfo{pages}{103009}\bibinfo{year}{ (\bibinfo{year}{2016})}}.

\bibitem[{\citenamefont{Hagmann and Bernstein}(2004)}]{HagmannBernstein}
\bibinfo{author}{\bibfnamefont{C.}~\bibnamefont{Hagmann}} \bibnamefont{and}
  \bibinfo{author}{\bibfnamefont{A.}~\bibnamefont{Bernstein}},
  \href{http://dx.doi.org/10.1109/TNS.2004.836061}{\bibinfo{journal}{IEEE
  Transactions on Nuclear Science}, \textbf{\bibinfo{volume}{51}},
  \bibinfo{pages}{2151}\bibinfo{year}{ (\bibinfo{year}{2004})}}, ISSN
  \bibinfo{issn}{0018-9499}.

\bibitem[{\citenamefont{Aprile et~al.}(2006)\citenamefont{Aprile, Dahl,
  de~Viveiros, Gaitskell, Giboni, Kwong, Majewski, Ni, Shutt, and
  Yamashita}}]{ColumbiaQy2006}
\bibinfo{author}{\bibfnamefont{E.}~\bibnamefont{Aprile}},
  \bibinfo{author}{\bibfnamefont{C.~E.} \bibnamefont{Dahl}},
  \bibinfo{author}{\bibfnamefont{L.}~\bibnamefont{de~Viveiros}},
  \bibinfo{author}{\bibfnamefont{R.~J.} \bibnamefont{Gaitskell}},
  \bibinfo{author}{\bibfnamefont{K.~L.} \bibnamefont{Giboni}},
  \bibinfo{author}{\bibfnamefont{J.}~\bibnamefont{Kwong}},
  \bibinfo{author}{\bibfnamefont{P.}~\bibnamefont{Majewski}},
  \bibinfo{author}{\bibfnamefont{K.}~\bibnamefont{Ni}},
  \bibinfo{author}{\bibfnamefont{T.}~\bibnamefont{Shutt}}, \bibnamefont{and}
  \bibinfo{author}{\bibfnamefont{M.}~\bibnamefont{Yamashita}},
  \href{http://dx.doi.org/10.1103/PhysRevLett.97.081302}{\bibinfo{journal}{Phys.
  Rev. Lett.}, \textbf{\bibinfo{volume}{97}},
  \bibinfo{pages}{081302}\bibinfo{year}{ (\bibinfo{year}{2006})}}.

\bibitem[{\citenamefont{Horn et~al.}(2011)\citenamefont{Horn, Belov, Akimov,
  Ara{\'u}jo, Barnes, Burenkov, Chepel, Currie, Edwards, Ghag
  et~al.}}]{HornQy2011}
\bibinfo{author}{\bibfnamefont{M.}~\bibnamefont{Horn}},
  \bibinfo{author}{\bibfnamefont{V.}~\bibnamefont{Belov}},
  \bibinfo{author}{\bibfnamefont{D.}~\bibnamefont{Akimov}},
  \bibinfo{author}{\bibfnamefont{H.}~\bibnamefont{Ara{\'u}jo}},
  \bibinfo{author}{\bibfnamefont{E.}~\bibnamefont{Barnes}},
  \bibinfo{author}{\bibfnamefont{A.}~\bibnamefont{Burenkov}},
  \bibinfo{author}{\bibfnamefont{V.}~\bibnamefont{Chepel}},
  \bibinfo{author}{\bibfnamefont{A.}~\bibnamefont{Currie}},
  \bibinfo{author}{\bibfnamefont{B.}~\bibnamefont{Edwards}},
  \bibinfo{author}{\bibfnamefont{C.}~\bibnamefont{Ghag}}, \bibnamefont{et~al.},
   \href{http://dx.doi.org/http://dx.doi.org/10.1016/j.physletb.2011.10.038}{\bibinfo{journal}{Physics
  Letters B}, \textbf{\bibinfo{volume}{705}}, \bibinfo{pages}{471
  }\bibinfo{year}{ (\bibinfo{year}{2011})}}, ISSN \bibinfo{issn}{0370-2693}.

\bibitem[{\citenamefont{Sorensen et~al.}(2009)\citenamefont{Sorensen, Manzur,
  Dahl, Angle, Aprile, Arneodo, Baudis, Bernstein, Bolozdynya, Coelho
  et~al.}}]{SorensenQy2009}
\bibinfo{author}{\bibfnamefont{P.}~\bibnamefont{Sorensen}},
  \bibinfo{author}{\bibfnamefont{A.}~\bibnamefont{Manzur}},
  \bibinfo{author}{\bibfnamefont{C.}~\bibnamefont{Dahl}},
  \bibinfo{author}{\bibfnamefont{J.}~\bibnamefont{Angle}},
  \bibinfo{author}{\bibfnamefont{E.}~\bibnamefont{Aprile}},
  \bibinfo{author}{\bibfnamefont{F.}~\bibnamefont{Arneodo}},
  \bibinfo{author}{\bibfnamefont{L.}~\bibnamefont{Baudis}},
  \bibinfo{author}{\bibfnamefont{A.}~\bibnamefont{Bernstein}},
  \bibinfo{author}{\bibfnamefont{A.}~\bibnamefont{Bolozdynya}},
  \bibinfo{author}{\bibfnamefont{L.}~\bibnamefont{Coelho}},
  \bibnamefont{et~al.},
  \href{http://dx.doi.org/http://dx.doi.org/10.1016/j.nima.2008.12.197}{\bibinfo{journal}{Nuclear
  Instruments and Methods in Physics Research Section A: Accelerators,
  Spectrometers, Detectors and Associated Equipment},
  \textbf{\bibinfo{volume}{601}}, \bibinfo{pages}{339 }\bibinfo{year}{
  (\bibinfo{year}{2009})}}, ISSN \bibinfo{issn}{0168-9002}.

\bibitem[{\citenamefont{Aprile et~al.}(2013)\citenamefont{Aprile, Alfonsi,
  Arisaka, Arneodo, Balan, Baudis, Bauermeister, Behrens, Beltrame, Bokeloh
  et~al.}}]{Xenon100Qy2013}
\bibinfo{author}{\bibfnamefont{E.}~\bibnamefont{Aprile}},
  \bibinfo{author}{\bibfnamefont{M.}~\bibnamefont{Alfonsi}},
  \bibinfo{author}{\bibfnamefont{K.}~\bibnamefont{Arisaka}},
  \bibinfo{author}{\bibfnamefont{F.}~\bibnamefont{Arneodo}},
  \bibinfo{author}{\bibfnamefont{C.}~\bibnamefont{Balan}},
  \bibinfo{author}{\bibfnamefont{L.}~\bibnamefont{Baudis}},
  \bibinfo{author}{\bibfnamefont{B.}~\bibnamefont{Bauermeister}},
  \bibinfo{author}{\bibfnamefont{A.}~\bibnamefont{Behrens}},
  \bibinfo{author}{\bibfnamefont{P.}~\bibnamefont{Beltrame}},
  \bibinfo{author}{\bibfnamefont{K.}~\bibnamefont{Bokeloh}},
  \bibnamefont{et~al.} (\bibinfo{collaboration}{XENON100 Collaboration}),
  \href{http://dx.doi.org/10.1103/PhysRevD.88.012006}{\bibinfo{journal}{Phys.
  Rev. D}, \textbf{\bibinfo{volume}{88}},
  \bibinfo{pages}{012006}\bibinfo{year}{ (\bibinfo{year}{2013})}}.

\bibitem[{\citenamefont{Manzur et~al.}(2010)\citenamefont{Manzur, Curioni,
  Kastens, McKinsey, Ni, and Wongjirad}}]{ManzurQy2010}
\bibinfo{author}{\bibfnamefont{A.}~\bibnamefont{Manzur}},
  \bibinfo{author}{\bibfnamefont{A.}~\bibnamefont{Curioni}},
  \bibinfo{author}{\bibfnamefont{L.}~\bibnamefont{Kastens}},
  \bibinfo{author}{\bibfnamefont{D.~N.} \bibnamefont{McKinsey}},
  \bibinfo{author}{\bibfnamefont{K.}~\bibnamefont{Ni}}, \bibnamefont{and}
  \bibinfo{author}{\bibfnamefont{T.}~\bibnamefont{Wongjirad}},
  \href{http://dx.doi.org/10.1103/PhysRevC.81.025808}{\bibinfo{journal}{Phys.
  Rev. C}, \textbf{\bibinfo{volume}{81}},
  \bibinfo{pages}{025808}\bibinfo{year}{ (\bibinfo{year}{2010})}}.

\bibitem[{\citenamefont{Akerib et~al.}()\citenamefont{Akerib, Ara{\'u}jo, Bai,
  Bailey, Balajthy, Beltrame, Bernard, Bernstein, Biesiadzinski, Boulton
  et~al.}}]{LUXDD}
\bibinfo{year}{ (????)}, \bibinfo{journal}{Submitted to Phys. Rev. C}.

\bibitem[{\citenamefont{Aprile et~al.}(2018{\natexlab{b}})\citenamefont{Aprile,
  Anthony, Lin, Greene, de~Perio, Gao, Howlett, Plante, Zhang, and
  Zhu}}]{Aprile2018_Yields}
\bibinfo{author}{\bibfnamefont{E.}~\bibnamefont{Aprile}},
  \bibinfo{author}{\bibfnamefont{M.}~\bibnamefont{Anthony}},
  \bibinfo{author}{\bibfnamefont{Q.}~\bibnamefont{Lin}},
  \bibinfo{author}{\bibfnamefont{Z.}~\bibnamefont{Greene}},
  \bibinfo{author}{\bibfnamefont{P.}~\bibnamefont{de~Perio}},
  \bibinfo{author}{\bibfnamefont{F.}~\bibnamefont{Gao}},
  \bibinfo{author}{\bibfnamefont{J.}~\bibnamefont{Howlett}},
  \bibinfo{author}{\bibfnamefont{G.}~\bibnamefont{Plante}},
  \bibinfo{author}{\bibfnamefont{Y.}~\bibnamefont{Zhang}}, \bibnamefont{and}
  \bibinfo{author}{\bibfnamefont{T.}~\bibnamefont{Zhu}},
  \href{http://dx.doi.org/10.1103/PhysRevD.98.112003}{\bibinfo{journal}{Phys.
  Rev. D}, \textbf{\bibinfo{volume}{98}},
  \bibinfo{pages}{112003}\bibinfo{year}{
  (\bibinfo{year}{2018}{\natexlab{b}})}}.

\bibitem[{\citenamefont{Edwards et~al.}(2018)\citenamefont{Edwards, Bernard,
  Boulton, Destefano, Gai, Horn, Larsen, Tennyson, Tvrznikova, Wahl
  et~al.}}]{PIXeY2018_EEE}
\bibinfo{author}{\bibfnamefont{B.}~\bibnamefont{Edwards}},
  \bibinfo{author}{\bibfnamefont{E.}~\bibnamefont{Bernard}},
  \bibinfo{author}{\bibfnamefont{E.}~\bibnamefont{Boulton}},
  \bibinfo{author}{\bibfnamefont{N.}~\bibnamefont{Destefano}},
  \bibinfo{author}{\bibfnamefont{M.}~\bibnamefont{Gai}},
  \bibinfo{author}{\bibfnamefont{M.}~\bibnamefont{Horn}},
  \bibinfo{author}{\bibfnamefont{N.}~\bibnamefont{Larsen}},
  \bibinfo{author}{\bibfnamefont{B.}~\bibnamefont{Tennyson}},
  \bibinfo{author}{\bibfnamefont{L.}~\bibnamefont{Tvrznikova}},
  \bibinfo{author}{\bibfnamefont{C.}~\bibnamefont{Wahl}}, \bibnamefont{et~al.},
   \href{http://stacks.iop.org/1748-0221/13/i=01/a=P01005}{\bibinfo{journal}{Journal
  of Instrumentation}, \textbf{\bibinfo{volume}{13}},
  \bibinfo{pages}{P01005}\bibinfo{year}{ (\bibinfo{year}{2018})}}.

\bibitem[{\citenamefont{Xu et~al.}(2019)\citenamefont{Xu, Pereverzev, Lenardo,
  Kingston, Naim, Bernstein, Kazkaz, and Tripathi}}]{LLNL_EEE2019}
\bibinfo{year}{ (\bibinfo{year}{2019})}, \bibinfo{journal}{In preparation}.

\bibitem[{\citenamefont{{Stephenson} et~al.}(2015)\citenamefont{{Stephenson},
  {Haefner}, {Lin}, {Ni}, {Pushkin}, {Raymond}, {Schubnell}, {Shutty},
  {Tarl{\'e}}, {Weaverdyck} et~al.}}]{StephensonMIXPaper}
\bibinfo{author}{\bibfnamefont{S.}~\bibnamefont{{Stephenson}}},
  \bibinfo{author}{\bibfnamefont{J.}~\bibnamefont{{Haefner}}},
  \bibinfo{author}{\bibfnamefont{Q.}~\bibnamefont{{Lin}}},
  \bibinfo{author}{\bibfnamefont{K.}~\bibnamefont{{Ni}}},
  \bibinfo{author}{\bibfnamefont{K.}~\bibnamefont{{Pushkin}}},
  \bibinfo{author}{\bibfnamefont{R.}~\bibnamefont{{Raymond}}},
  \bibinfo{author}{\bibfnamefont{M.}~\bibnamefont{{Schubnell}}},
  \bibinfo{author}{\bibfnamefont{N.}~\bibnamefont{{Shutty}}},
  \bibinfo{author}{\bibfnamefont{G.}~\bibnamefont{{Tarl{\'e}}}},
  \bibinfo{author}{\bibfnamefont{C.}~\bibnamefont{{Weaverdyck}}},
  \bibnamefont{et~al.},
  \href{http://arxiv.org/abs/1507.01310}{\bibinfo{journal}{ArXiv
  e-prints}:1507.01310\bibinfo{year}{ (\bibinfo{year}{2015})}}.

\bibitem[{\citenamefont{Akerib et~al.}(2012)\citenamefont{Akerib, Bai,
  Bedikian, Bernard, Bernstein, Bradley, Cahn, Carmona-Benitez, Carr, Chapman
  et~al.}}]{LUXSim_Paper}
\bibinfo{author}{\bibfnamefont{D.}~\bibnamefont{Akerib}},
  \bibinfo{author}{\bibfnamefont{X.}~\bibnamefont{Bai}},
  \bibinfo{author}{\bibfnamefont{S.}~\bibnamefont{Bedikian}},
  \bibinfo{author}{\bibfnamefont{E.}~\bibnamefont{Bernard}},
  \bibinfo{author}{\bibfnamefont{A.}~\bibnamefont{Bernstein}},
  \bibinfo{author}{\bibfnamefont{A.}~\bibnamefont{Bradley}},
  \bibinfo{author}{\bibfnamefont{S.}~\bibnamefont{Cahn}},
  \bibinfo{author}{\bibfnamefont{M.}~\bibnamefont{Carmona-Benitez}},
  \bibinfo{author}{\bibfnamefont{D.}~\bibnamefont{Carr}},
  \bibinfo{author}{\bibfnamefont{J.}~\bibnamefont{Chapman}},
  \bibnamefont{et~al.},
  \href{http://dx.doi.org/http://dx.doi.org/10.1016/j.nima.2012.02.010}{\bibinfo{journal}{Nucl.
  Instr. Meth. Phys. Res. A}, \textbf{\bibinfo{volume}{675}},
  \bibinfo{pages}{63 }\bibinfo{year}{ (\bibinfo{year}{2012})}}, ISSN
  \bibinfo{issn}{0168-9002}.

\bibitem[{\citenamefont{Doke et~al.}(1976)\citenamefont{Doke, Hitachi, Kubota,
  Nakamoto, and Takahashi}}]{DokeFanoFactors}
\bibinfo{author}{\bibfnamefont{T.}~\bibnamefont{Doke}},
  \bibinfo{author}{\bibfnamefont{A.}~\bibnamefont{Hitachi}},
  \bibinfo{author}{\bibfnamefont{S.}~\bibnamefont{Kubota}},
  \bibinfo{author}{\bibfnamefont{A.}~\bibnamefont{Nakamoto}}, \bibnamefont{and}
  \bibinfo{author}{\bibfnamefont{T.}~\bibnamefont{Takahashi}},
  \href{http://dx.doi.org/http://dx.doi.org/10.1016/0029-554X(76)90292-5}{\bibinfo{journal}{Nuclear
  Instruments and Methods}, \textbf{\bibinfo{volume}{134}}, \bibinfo{pages}{353
  }\bibinfo{year}{ (\bibinfo{year}{1976})}}, ISSN \bibinfo{issn}{0029-554X}.

\bibitem[{\citenamefont{Akerib et~al.}(2017{\natexlab{b}})\citenamefont{Akerib,
  Alsum, Ara\'ujo, Bai, Bailey, Balajthy, Beltrame, Bernard, Bernstein,
  Biesiadzinski et~al.}}]{LUX2017_Yield}
\bibinfo{author}{\bibfnamefont{D.~S.} \bibnamefont{Akerib}},
  \bibinfo{author}{\bibfnamefont{S.}~\bibnamefont{Alsum}},
  \bibinfo{author}{\bibfnamefont{H.~M.} \bibnamefont{Ara\'ujo}},
  \bibinfo{author}{\bibfnamefont{X.}~\bibnamefont{Bai}},
  \bibinfo{author}{\bibfnamefont{A.~J.} \bibnamefont{Bailey}},
  \bibinfo{author}{\bibfnamefont{J.}~\bibnamefont{Balajthy}},
  \bibinfo{author}{\bibfnamefont{P.}~\bibnamefont{Beltrame}},
  \bibinfo{author}{\bibfnamefont{E.~P.} \bibnamefont{Bernard}},
  \bibinfo{author}{\bibfnamefont{A.}~\bibnamefont{Bernstein}},
  \bibinfo{author}{\bibfnamefont{T.~P.} \bibnamefont{Biesiadzinski}},
  \bibnamefont{et~al.} (\bibinfo{collaboration}{LUX Collaboration}),
  \href{http://dx.doi.org/10.1103/PhysRevD.95.012008}{\bibinfo{journal}{Phys.
  Rev. D}, \textbf{\bibinfo{volume}{95}},
  \bibinfo{pages}{012008}\bibinfo{year}{
  (\bibinfo{year}{2017}{\natexlab{b}})}}.

\bibitem[{\citenamefont{Akerib et~al.}(2016{\natexlab{b}})\citenamefont{Akerib,
  Ara\'ujo, Bai, Bailey, Balajthy, Beltrame, Bernard, Bernstein, Biesiadzinski,
  Boulton et~al.}}]{LUXTritium}
\bibinfo{author}{\bibfnamefont{D.~S.} \bibnamefont{Akerib}},
  \bibinfo{author}{\bibfnamefont{H.~M.} \bibnamefont{Ara\'ujo}},
  \bibinfo{author}{\bibfnamefont{X.}~\bibnamefont{Bai}},
  \bibinfo{author}{\bibfnamefont{A.~J.} \bibnamefont{Bailey}},
  \bibinfo{author}{\bibfnamefont{J.}~\bibnamefont{Balajthy}},
  \bibinfo{author}{\bibfnamefont{P.}~\bibnamefont{Beltrame}},
  \bibinfo{author}{\bibfnamefont{E.~P.} \bibnamefont{Bernard}},
  \bibinfo{author}{\bibfnamefont{A.}~\bibnamefont{Bernstein}},
  \bibinfo{author}{\bibfnamefont{T.~P.} \bibnamefont{Biesiadzinski}},
  \bibinfo{author}{\bibfnamefont{E.~M.} \bibnamefont{Boulton}},
  \bibnamefont{et~al.} (\bibinfo{collaboration}{LUX Collaboration}),
  \href{http://dx.doi.org/10.1103/PhysRevD.93.072009}{\bibinfo{journal}{Phys.
  Rev. D}, \textbf{\bibinfo{volume}{93}},
  \bibinfo{pages}{072009}\bibinfo{year}{
  (\bibinfo{year}{2016}{\natexlab{b}})}}.

\bibitem[{\citenamefont{Albert et~al.}(2017)\citenamefont{Albert, Barbeau,
  Beck, Belov, Breidenbach, Brunner, Burenkov, Cao, Cen, Chambers
  et~al.}}]{EXO2017_EDrift}
\bibinfo{author}{\bibfnamefont{J.~B.} \bibnamefont{Albert}},
  \bibinfo{author}{\bibfnamefont{P.~S.} \bibnamefont{Barbeau}},
  \bibinfo{author}{\bibfnamefont{D.}~\bibnamefont{Beck}},
  \bibinfo{author}{\bibfnamefont{V.}~\bibnamefont{Belov}},
  \bibinfo{author}{\bibfnamefont{M.}~\bibnamefont{Breidenbach}},
  \bibinfo{author}{\bibfnamefont{T.}~\bibnamefont{Brunner}},
  \bibinfo{author}{\bibfnamefont{A.}~\bibnamefont{Burenkov}},
  \bibinfo{author}{\bibfnamefont{G.~F.} \bibnamefont{Cao}},
  \bibinfo{author}{\bibfnamefont{W.~R.} \bibnamefont{Cen}},
  \bibinfo{author}{\bibfnamefont{C.}~\bibnamefont{Chambers}},
  \bibnamefont{et~al.} (\bibinfo{collaboration}{EXO-200 Collaboration}),
  \href{http://dx.doi.org/10.1103/PhysRevC.95.025502}{\bibinfo{journal}{Phys.
  Rev. C}, \textbf{\bibinfo{volume}{95}},
  \bibinfo{pages}{025502}\bibinfo{year}{ (\bibinfo{year}{2017})}}.

\bibitem[{\citenamefont{Baker and
  Cousins}(1984)}]{BakerCousins1983_SaturatedModels}
\bibinfo{author}{\bibfnamefont{S.}~\bibnamefont{Baker}} \bibnamefont{and}
  \bibinfo{author}{\bibfnamefont{R.~D.} \bibnamefont{Cousins}},
  \href{http://dx.doi.org/https://doi.org/10.1016/0167-5087(84)90016-4}{\bibinfo{journal}{Nuclear
  Instruments and Methods in Physics Research}, \textbf{\bibinfo{volume}{221}},
  \bibinfo{pages}{437 }\bibinfo{year}{ (\bibinfo{year}{1984})}}, ISSN
  \bibinfo{issn}{0167-5087}.

\bibitem[{\citenamefont{Szydagis et~al.}(2011)\citenamefont{Szydagis, Barry,
  Kazkaz, Mock, Stolp, Sweany, Tripathi, Uvarov, Walsh, and Woods}}]{NESTpaper}
\bibinfo{author}{\bibfnamefont{M.}~\bibnamefont{Szydagis}},
  \bibinfo{author}{\bibfnamefont{N.}~\bibnamefont{Barry}},
  \bibinfo{author}{\bibfnamefont{K.}~\bibnamefont{Kazkaz}},
  \bibinfo{author}{\bibfnamefont{J.}~\bibnamefont{Mock}},
  \bibinfo{author}{\bibfnamefont{D.}~\bibnamefont{Stolp}},
  \bibinfo{author}{\bibfnamefont{M.}~\bibnamefont{Sweany}},
  \bibinfo{author}{\bibfnamefont{M.}~\bibnamefont{Tripathi}},
  \bibinfo{author}{\bibfnamefont{S.}~\bibnamefont{Uvarov}},
  \bibinfo{author}{\bibfnamefont{N.}~\bibnamefont{Walsh}}, \bibnamefont{and}
  \bibinfo{author}{\bibfnamefont{M.}~\bibnamefont{Woods}},
  \href{http://dx.doi.org/10.1088/1748-0221/6/10/P10002}{\bibinfo{journal}{J.
  Instrum.}, \textbf{\bibinfo{volume}{6}},
  \bibinfo{pages}{P10002}\bibinfo{year}{ (\bibinfo{year}{2011})}}.

\bibitem[{\citenamefont{Szydagis et~al.}(2013)\citenamefont{Szydagis, Fyhrie,
  Thorngren, and Tripathi}}]{NESTpaper2}
\bibinfo{author}{\bibfnamefont{M.}~\bibnamefont{Szydagis}},
  \bibinfo{author}{\bibfnamefont{A.}~\bibnamefont{Fyhrie}},
  \bibinfo{author}{\bibfnamefont{D.}~\bibnamefont{Thorngren}},
  \bibnamefont{and} \bibinfo{author}{\bibfnamefont{M.}~\bibnamefont{Tripathi}},
   \href{http://dx.doi.org/10.1088/1748-0221/8/10/C10003}{\bibinfo{journal}{J.
  Instrum.}, \textbf{\bibinfo{volume}{8}},
  \bibinfo{pages}{C10003}\bibinfo{year}{ (\bibinfo{year}{2013})}}.

\bibitem[{\citenamefont{Szydagis et~al.}(2018)\citenamefont{Szydagis, Balajthy,
  Brodsky, Cutter, Huang, Kozlova, Lenardo, Manalaysay, McKinsey, Mooney
  et~al.}}]{NESTZenodo}
\bibinfo{author}{\bibfnamefont{M.}~\bibnamefont{Szydagis}},
  \bibinfo{author}{\bibfnamefont{J.}~\bibnamefont{Balajthy}},
  \bibinfo{author}{\bibfnamefont{J.}~\bibnamefont{Brodsky}},
  \bibinfo{author}{\bibfnamefont{J.}~\bibnamefont{Cutter}},
  \bibinfo{author}{\bibfnamefont{J.}~\bibnamefont{Huang}},
  \bibinfo{author}{\bibfnamefont{E.}~\bibnamefont{Kozlova}},
  \bibinfo{author}{\bibfnamefont{B.}~\bibnamefont{Lenardo}},
  \bibinfo{author}{\bibfnamefont{A.}~\bibnamefont{Manalaysay}},
  \bibinfo{author}{\bibfnamefont{D.}~\bibnamefont{McKinsey}},
  \bibinfo{author}{\bibfnamefont{M.}~\bibnamefont{Mooney}},
  \bibnamefont{et~al.}, \emph{\bibinfo{title}{{Noble Element Simulation
  Technique v2.0}}} (\bibinfo{year}{2018}),
  \urlprefix\url{https://doi.org/10.5281/zenodo.1314669}.

\end{thebibliography}
